\documentclass[aps,twocolumn,showpacs,amsmath,amssymb,superscriptaddress,longbibliography,prb]{revtex4-2}
\usepackage[colorlinks=true,citecolor=blue,linkcolor=blue,hypertexnames=false]{hyperref}
\usepackage{amsmath}
\usepackage{dsfont}
\usepackage{hyperref}
\usepackage{textcomp}
\usepackage{tabularx}
\usepackage{graphicx,tikz}
\usepackage{amsfonts}

\newcommand{\bxi}{\boldsymbol{\xi}}

\def \ee{\end{equation}}
\def \bea{\begin{eqnarray}}
\def \eea{\end{eqnarray}}

\def \bea{\begin{eqnarray}}
\def \eea{\end{eqnarray}}

\begin{document}
\title{An integrated neural wavefunction solver for spinful Fermi systems}

\author{Alexander Avdoshkin}
\affiliation{Department of Physics, Massachusetts Institute of Technology, Cambridge, MA 02139, USA}
\author{Max Geier}
\affiliation{Department of Physics, Massachusetts Institute of Technology, Cambridge, MA 02139, USA}
\author{Liang Fu}
\affiliation{Department of Physics, Massachusetts Institute of Technology, Cambridge, MA 02139, USA}

%
%
%
%

\begin{abstract}
    %
    We present an approach to solving the ground state of Fermi systems that contain spin or other discrete degrees of freedom in addition to continuous coordinates. 
    The approach combines a Markov chain Monte Carlo sampling for energy estimation that we adapted to cover the extended configuration space with a transformer-based wavefunction to represent fermionic states. 
    A transformer with both continuous position and discrete spin as inputs achieves universal approximation to spinful generalized orbitals.
    We validate the method on a range of two-dimensional material problems: a two-dimensional electron gas with Rashba spin-orbit coupling, a noncollinear spin texture, and a quantum antiferromagnet in a honeycomb moir\'e potential.
\end{abstract}

\maketitle

\section{Introduction}
Determining ground-state wave functions of interacting quantum many-body systems is a central problem in quantum chemistry, cold atom, and condensed matter physics. 
The electronic many-body state underlies phenomena ranging from 
magnetism and superconductivity to fractionalization in topological phases. 
Its accurate computation is challenging because of quantum correlations between many constituent particles. 


Recently, neural network variational Monte Carlo (NN-VMC)     \cite{Carleo2017,pfau2020ab,Hermann2020PauliNet,von2022self} has emerged as a powerful tool to determine ground states of continuum quantum systems with high precision \cite{pescia2022neural,choo2020fermionic}. It has succeeded in achieving chemical accuracy \cite{jiang2025, foster2025ab} in molecular systems, and in solving strongly correlated electron problems in quantum matter including Wigner crystals \cite{cassella2023discovering, pescia2024message}, chiral superconductivity \cite{li2025attention}, fractional quantum Hall liquids \cite{teng2024solving, qian2025describing,nazaryan2025artificial}  and fractional Chern insulators \cite{luo2025solving, li2025deep}. 
NN-VMC has also been extensively applied to lattice systems \cite{carleo2019netket} such as fermionic Hubbard model \cite{gu2025solving,luo2019backflow,robledo2022fermionic}  and Heisenberg spin model \cite{PhysRevB.108.054410, astrakhantsev2021broken,choo2019two}. 


%


Problems that involve both continuous positions and discrete spin (or pseudospin) degrees of freedom (DOFs) are much less studied \cite{luo2025solving, adams2021variational, li2025deep, zhan2025}.
Yet, in quantum matter, the interplay between spin and position DOFs leads to a variety of correlated effects such as antiferromagnitism \cite{spintronics_baltz}, multiferroics \cite{multiferroics_spaldin} and skyrmions \cite{skyrmions_nagaosa}. 
Similarly, spin-momentum coupling, arising from spin-orbit interaction, can produce topologically non-trivial band structures leading to topological insulators \cite{moore2010birth, fu2007topological,hasan2010colloquium} and Weyl/Dirac semimetals \cite{armitage2018weyl}.

The simultaneous presence of continuous and discrete variables provides a challenge both in choosing an appropriate wavefunction ansatz and in ensuring efficient energy estimation from the Monte Carlo sampler which is necessary for reliable optimization. In constructing the architecture one needs to universally express correlations between electronic positions as well as their spin variables while respecting the Pauli antisymmetry principle. The Monte Carlo sampler needs to perform updates of both types of DOFs to efficiently cover the relevant part of the configuration space \cite{Melton2016QMCspin,Melton2016spinorbit,adams2021variational,gerard2024solids,luo2025solving,li2025deep,zhan2025}.

\begin{figure}
    \centering 
    \begin{tabular}{@{}l@{}}
        \textbf{(a)}\\
        ~~~~~~~\includegraphics[width=0.85\linewidth]{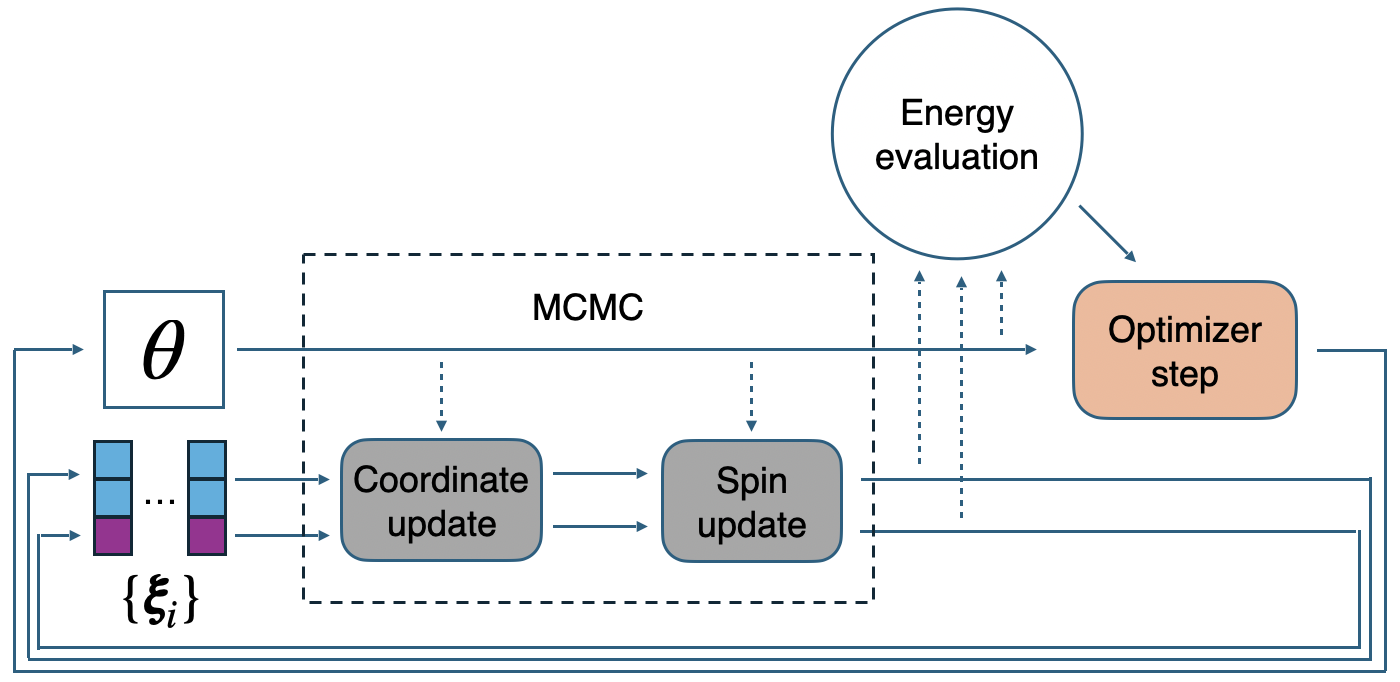}\\[0.5ex]
        \textbf{(b)}\\
        \includegraphics[width=.95\linewidth]{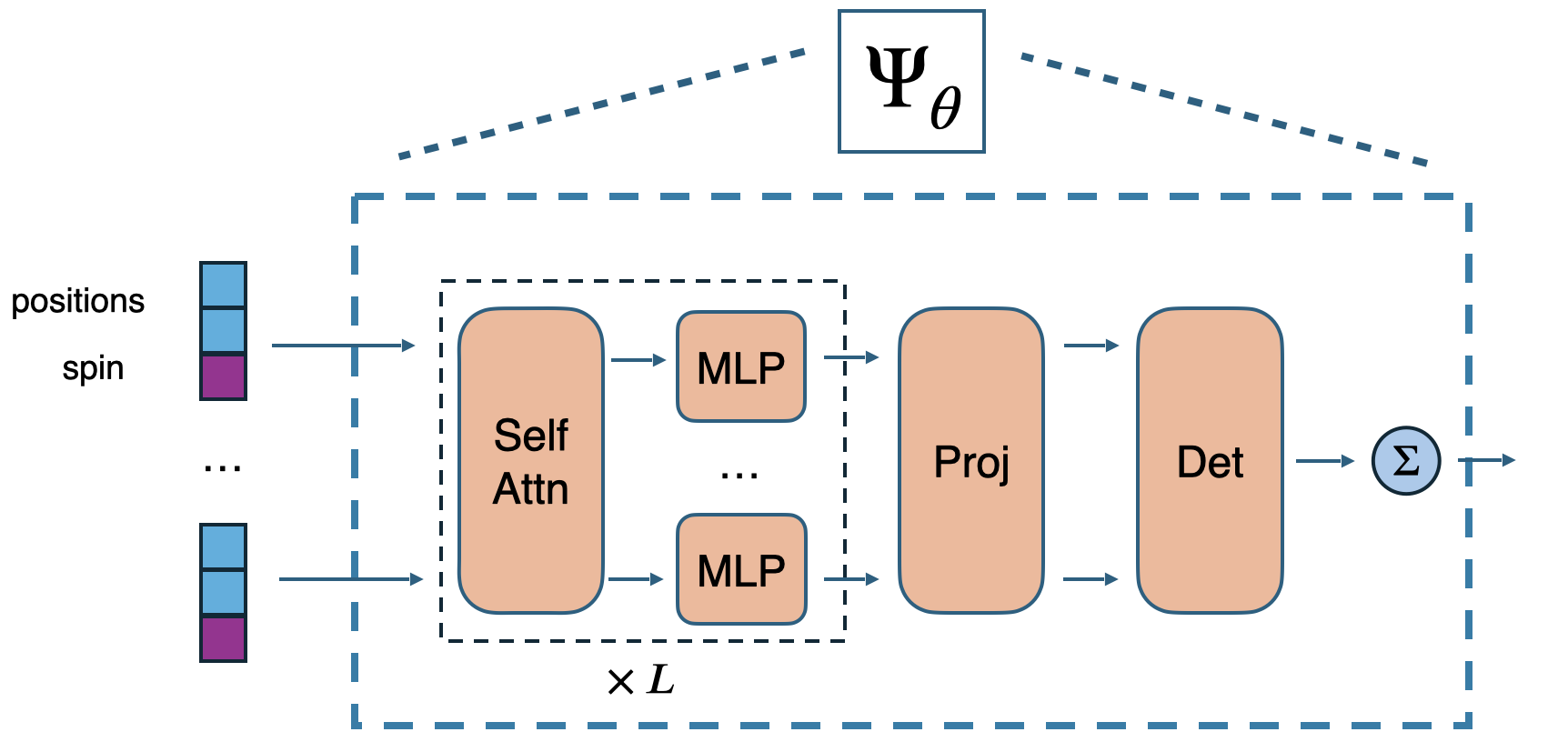}
    \end{tabular}
    \caption{(a) Schematic representation of the VMC training loop. The wavefunction parameters $\theta$ and the batch configurations $\{\bxi_i\}$ are updated by a series of consecutive Monte Carlo sampling steps that modify the batch and optimizer steps that modify the wavefunction parameters. 
    (b) Architecture of the neural network representing the wavefunction $\Psi_{\theta}$. Particle position and spin (or discrete DOFs) are processed in streams that affect each other through the attention mechanism.
    }
    \label{fig:scheme}
\end{figure}

In this work, we solve both challenges: first, we achieve efficient sampling by including separate spin updates in the Monte Carlo routine [Fig.~\ref{fig:scheme}(a)].
Second, we establish that processing both continuous position and discrete spin jointly provides a universal approximator of generalized orbitals 
employed in constructing fermionic wavefunctions \cite{pfau2020ab}.

We implement this approach using a transformer neural network \cite{vaswani2017attention} with determinants to enforce the Pauli principle \cite{von2022self} [see Fig.~\ref{fig:scheme}(b)]. Previously, this ansatz was able to describe metals \cite{geier2025attention}, charge-ordered insulators \cite{geier2025attention}, superconductors \cite{li2025attentionneedsolvechiral}, topological phases \cite{teng2024solving} without prior knowledge of the system.

We demonstrate that this approach is able to solve three different classes of spinful Fermi systems: (i) a Rashba spin-orbit coupled Fermi gas, (ii) a Fermi gas with non-collinear spin texture winding in real space, (iii) an antiferromagnet emerging from Coulomb repulsion in a two-dimensional electron gas with honeycomb potential.

This article is organized as follows. Sec. \ref{sec:VMC} reviews the VMC method. Sec. \ref{sec:mcmc} describes the MCMC procedure including both coordinate and spin updates. Sec. \ref{sec:neural_arch} describes the  attention-based wave function ansatz that can represent the most general spin-coordinate wavefunction and Sec. \ref{sec:local_energies} discusses the evaluation of local energies with spinful terms.  In Sec. \ref{sec:results}, we show simulation results and Sec. \ref{sec:conclusions} contains conclusions and outlook.   

\section{Variational Monte Carlo} \label{sec:VMC}



As other variational approaches, VMC uses an ansatz to approximate the desired quantum state. For an (unnormalized) wavefunction $\Psi_{\theta}$, where $\theta$ denotes the parameters, the ground state is identified by minimizing the expected energy:
\bea\label{eq:exp_energy}
E_{\theta} = \frac{\sum_{\{\boldsymbol{\xi}\},\{\boldsymbol{\xi'}\}}\Psi_{\theta}^*(\{\boldsymbol{\xi}\}) \hat{H}(\{\boldsymbol{\xi}\},\{\boldsymbol{\xi'}\}) \Psi_{\theta}(\{\boldsymbol{\xi'}\})}{\sum_{\{\boldsymbol{\xi}\}}| \Psi_{\theta}(\{\boldsymbol{\xi}\})|^2},
\eea
where $ \{\boldsymbol{\xi}\}$ stands for the configuration of the system (i.e. positions and spins of all particles) and $\hat{H}(\{\boldsymbol{\xi}\},\{\boldsymbol{\xi'}\})$ are the matrix elements of the Hamiltonian.

For many-body systems, direct integration in Eq. \eqref{eq:exp_energy} becomes prohibitively expensive computationally. In VMC, the integral is efficiently evaluated by sampling electron configurations $\boldsymbol{\Xi}$ from the wavefunction intensity $|\Psi(\boldsymbol{\Xi})|^2$,
\begin{align}
   E_{\theta} = & \  \mathbb{E}_{\{\boldsymbol{\xi}\} \sim |\Psi_{\theta}(\{\boldsymbol{\xi}\})|^2}[E_{{\rm loc},\theta}(\{\boldsymbol{\xi}\})],\label{eq:vmc}\\
E_{{\rm loc},\theta}(\{\boldsymbol{\xi}\}) = & \sum_{\{\boldsymbol{\xi'}\}}\Psi_{\theta}^{-1}(\{\boldsymbol{\xi}\}) \hat{H}(\{\boldsymbol{\xi}\},\{\boldsymbol{\xi'}\}) \Psi_{\theta}(\{\boldsymbol{\xi'}\})\label{eq:local_energy}, 
\end{align}
where the contributions of each configuration $E_{{\rm loc},\theta}(\{\boldsymbol{\xi'}\})$ are called local energies.

To make practical use of Eqs. \eqref{eq:vmc} and \eqref{eq:local_energy}, we need three ingredients: sampling of configurations according to $|\Psi_{\theta}(\{\boldsymbol{\xi}\})|^2$, numerically exact evaluation of local energies and a sufficiently expressive ansatz $\Psi_{\theta}$.

As we discuss in Sec. \ref{sec:mcmc}, the sampling in Eq. \eqref{eq:vmc} can be efficiently performed via a Markov chain Monte Carlo procedure that sequentially generates configurations $\{\boldsymbol{\xi}\}$ that follow the probability distribution set by $|\Psi_{\theta}(\{\boldsymbol{\xi}\})|^2$. 
We describe evaluation of local energies from Eq. \eqref{eq:local_energy} in Sec. \ref{sec:local_energies}.

The sampled energy expectation values and their gradients with respect to parameters are used to optimize the variational wavefunction towards representing the ground state by energy minimization. In practice, optimization strategies that approximate natural gradient descent \cite{Amari1998NGD}, such as stochastic reconfiguration \cite{Sorella1998,Stokes2020quantumnatural} and neural-network specific approximations for improved efficiency \cite{Martens2015KFAC,pfau2020ab,Chen2024minSR,Goldshlager2024SPRING,gu2025solvinghubbardmodelneural} perform best because they include information on the wavefunction geometry in parameter space. In this work, we employ Kronecker-factored approximate curvature \cite{Martens2015KFAC,pfau2020ab,geier2025attention} as optimizer, which includes an approximation to the Fisher-information matrix describing the geometry associated to the wavefunction intensity.

Finally, for a good performance of VMC, one needs to choose an ansatz $\Psi_{\theta}$ capable of giving a good approximation to the wavefunction of interest. In subsection \ref{sec:neural_arch}, we explain how transformers can be used to give a fermionic wavefunction that includes arbitrary DOFs.

\subsection{Spinful MCMC procedure} \label{sec:mcmc}

We use a generalized MCMC procedure which combines coordinate updates with spin and other discrete DOF updates. Each configuration of the system with $N$ particles is represented by a combined state variable $\{\bxi_i\}$. 

The procedure consists of proposing updates $\{\bxi_i\} \to \{\bxi'_i\}$ and accepting them with probability
\bea
\begin{cases}
|\psi'/\psi|^2, |\psi'| < |\psi|\\
1, |\psi'| \geq |\psi|,
\end{cases}
\eea
where $\psi = \psi(\{\bxi_i\})$ and $\psi' = \psi(\{\bxi'_i\})$ are the amplitudes of the wavefunctions evaluated for the old and new configurations, respectively.


Continuous and discrete updates are proposed and accepted separately, as is illustrated in Fig.~\ref{fig:scheme}(a). The proposal of real-space coordinate updates follows Gaussian probability:
\bea 
w(\bold{r}_i \to \bold{r}_i')  = \frac{1}{2 \pi \sigma^2} e^{-|\bold{r}_{i}-\bold{r}'_i|^2/2 \sigma^2},
\eea
with $\sigma$ being the width of the proposal.

When considering flip updates that change the total magnetization $s_z$, we consider a MCMC move proposal that randomly flips spins $\sigma_i \to - \sigma_i$ with uniform probability $p$ for each fermion, 
\bea
w(\sigma_i \to \sigma_i') =  p, ~ \sigma_i \neq \sigma_i'.
\eea
A similar procedure generalizes to arbitrary pseudospin $\alpha_i$.

For the cases where the Hamiltonian preserves magnetization $s_z$ quantization and it is preferable to remain in a fixed $s_z$ sector, we implemented \textit{sector-preserving updates} that permute the spin configuration without changing total magnetization: At each step, we first decide randomly the number $m$ of electron pairs whose spin states are going to be swapped. This number $m$ is drawn from a Poisson distribution with mean $\lambda = p_{\rm swap} N_e / 2$ where $N_e$ is the total number of electrons and $p_{\rm swap}$ a parameter. 
Then, $m$ spin swaps are performed by iterating the following procedure $m$ times: (i) Randomly and uniformly select a pair of electrons, then (ii) exchange their spin states.
The acceptance of spin updates is decided separately from coordinate updates. 
The advantage of this procedure is demonstrated in Sec. \ref{sec:afm}.

\subsection{Neural architecture} \label{sec:neural_arch}

Let us consider $N$ particles with positions $\bold{r}_i$ and spin/discrete DOFs $\alpha_i$ (when the model includes spin ($s_i$) and sublattice ($\tau_i$) DOFs we will have $\alpha_i = \{s_i, \tau_i\}$). Our goal is to be able to represent a many-particle wavefunction:
\bea\label{eq:wave_fun}
\Psi(\{\bold{r}_1,\alpha_1\},\{\bold{r}_2,\alpha_2\},\dots,\{\bold{r}_N,\alpha_N\})
\eea
that is anti-symmetric in the permutation of any pair of electrons $\{\bold{r}_i,\alpha_i\}\leftrightarrow
\{\bold{r}_j,\alpha_j\}$. In what follows, we introduce generalized coordinates $\boldsymbol{\xi}_i=\{\bold{r}_i,\alpha_i\}$ . 

A common approach is to start with a Slater determinant $\det\limits_{ij} \left[\phi_j(\bxi_i)\right]$ and then promote the single particle orbitals to generalized orbitals $\phi_j(\bxi_i, \bxi_{/i})$ \cite{pfau2020ab,Hermann2020PauliNet,von2022self}, where $\bxi_{/i}$ stands for all coordinates other than $\bxi_{i}$ with the dependence on those coordinates being permutation invariant. The generalized orbitals are motivated by the idea of {\it backflow} \cite{luo2019backflow,kwon1993backflow,Feynman1956} to capture electron correlations: The state of each electron is influenced by the states of all other electrons.

To achieve maximum expressive power, it is important to allow the most general $\phi_j(\bxi_i, \bxi_{/i})$. A common choice has been $\phi_j(\bxi_i, {\bf r}_{/i})$ \cite{li2025deep, zhan2025}, which allows for the most general position, but not spin, dependence. In Ref. \cite{zhan2025}, this type of ansatz was dubbed "spinor" as opposed to "generalized spinor" $\phi_j(\bxi_i, \bxi_{/i})$. 
%
%
In our approach, the fully expressive generalized orbitals $\phi_j(\bxi_i, \bxi_{/i})$ are obtained from a neural network that \emph{universally} approximates permutation equivariant sequence-to-sequence functions, when {\it all} particle DOFs -- including both position and spin -- are passed as inputs in a single vector: 
\bea \label{eq:coordinates}
\bold{l}_i = \begin{pmatrix}
r^x_i\\
r^y_i\\
s_i
\end{pmatrix}.
\eea
Other DOFs, if present, should be concatenated to this vector. 

The fermionic wavefunction is a sum of determinants of generalized orbitals:
\bea\label{eq:generalized_determinant_wf}
\Psi(\bxi_1,\dots,\bxi_N) = \sum_{m=1}^{N_{\rm det}} \det_{ij} \left[\phi^m_j(\bxi_i,\{\bxi_{/i}\})\right],
\eea
where $\{\bxi_{/i}\}$ denotes all generalized coordinates distinct from $i$.
Due to the structure of the attention mechanism, the generalized orbitals $\phi^m_j(\bxi_i,\{\bxi_{/i}\})$ are symmetric in the exchange of $\bxi_{/i}$. Taking the determinant at the end leads to an antisymmetric fermionic wavefunction. 
In practice, one constructs multiple sets of generalized orbitals from the same transformer NN and the final wavefunction is a sum of Slater determinants \cite{pfau2020ab,von2022self,geier2025attention}.

Ref.~\cite{Yun2020Are} established that transformers are universal approximators of continuous permutation equivariant sequence-to-sequence functions. 
Because the correlated orbitals $\phi^m_j(\bxi_i,\{\bxi_{/i}\})$ are obtained directly by projection from the transformer output, the transformer universally approximates correlated orbitals. As a consequence, any fermionic wavefunction that can be expressed as a sum of determinants of continuous, correlated orbitals, as in Eq.~\eqref{eq:generalized_determinant_wf}, is universally approximated by the transformer wavefunction architecture. 

%
%



\subsection{Evaluating local energies} \label{sec:local_energies}

We assume a Hamiltonian of the form
\bea\label{eq:Hamiltonian}
H = \sum_i \frac{\boldsymbol{\nabla}_i^2}{2m} + V(\bold{r}) + H_{\rm spin},
\eea
where $V(\bold{r})$ includes both single particle and interaction terms and $H_{\rm spin}$ will be introduced  below. 

Evaluating the local energies according to Eq. \eqref{eq:local_energy} becomes harder with the increasing number of off-diagonal matrix elements of $H$. Fortunately, the potential term is diagonal in the position and for the kinetic term reduces to taking the Laplacian of the wavefunction\cite{Foulkes_review} which can be efficiently computed for neural networks by using derivative propagation \cite{li2023forward}.
 
As concrete examples, we study two model Hamiltonians with spin-dependent terms. The first one is the spatially varying Zeeman term
\bea
H_{\rm spin} = \sum_i\bold{B}(\bold{r}_i)\cdot \boldsymbol{\sigma}_i,
\eea
which we use in Sec. \ref{sec:spin-spiral}.
The corresponding local energy is evaluated as 
\bea \label{eq:term_zeeman}
E^{\rm (local)}_{\rm spin} = \sum_{i,\mu,\alpha'} B^{\mu}(\bold{r}_i) \Psi(\bold{r}_i,\alpha_i)^{-1} \sigma^{\mu}_{\alpha_i\alpha'}\Psi(\bold{r}_i,\alpha'),
\eea
where $\sigma^{\lambda}_{\alpha\alpha'}$ denote the matrix elements of Pauli matrices $\sigma^{\lambda}$. 

The second possibility is the spin-orbit coupling
\bea
H_{\rm spin} = \sum_{i,\mu,\nu}\kappa_{\mu\nu} \hat{p}^{\nu}_i\sigma^{\mu}_i,
\eea
where $\kappa_{\mu\nu}$ specifies the spin-orbit interaction, i.e. $\kappa = \{\{0,-1\},\{1,0\}\}$ for the Rashba coupling (see Sec. \ref{sec:soc}). In this case, the local energy takes the form
\bea \label{eq:term_soc}
E^{\rm (local)}_{\rm spin} = \sum_{i, \mu,\nu,\alpha'} \kappa_{\mu 
\nu} \Psi(\bold{r}_i,\alpha_i)^{-1} \sigma^{\mu}_{\alpha_i\alpha'}\partial_{\nu}\Psi(\bold{r}_i,\alpha').
\eea
The presence of the first derivative in Eq. \eqref{eq:term_soc} makes it computationally expensive compared to Eq. \eqref{eq:term_zeeman}, but it remains cheap compared to the Laplacian computation in the kinetic energy.

\section{Simulation results} \label{sec:results}

Here we demonstrate the ability of our method to accurately capture the ground states of the various model systems. The hyperparameters for the simulations are delegated to App. \ref{app:hyperparameters}.

\subsection{Spin spiral} \label{sec:spin-spiral}

As a simple model with nontrivial spin dependence of the Hamiltonian we consider the spin spiral Hamiltonian \cite{onishi2025emergent}:
\bea\label{eq:spin_spiral}
H = {\bf p}^2/2 - J {\bf S}(\bold{r}) \cdot \boldsymbol{\sigma},
\eea
with a particular choice ${\bf S}(\bold{r}) = (\cos({\bf q}\cdot {\bf r}),\sin({\bf q}\cdot {\bf r}),0)$. The exact spectrum of this Hamiltonian (lower branch) is $E({\bf p}) = ({\bf p}^2 + {\bf q}^2/4)/2 - \sqrt{J^2 + ({\bf q }\cdot{\bf p })^2/4}$.
Eq. \eqref{eq:spin_spiral} lies in the same class of models as the effective model of twisted MoTe$_2$ \cite{luo2025solving}. 

\begin{figure}
    \centering
    \includegraphics[width=1.\linewidth]{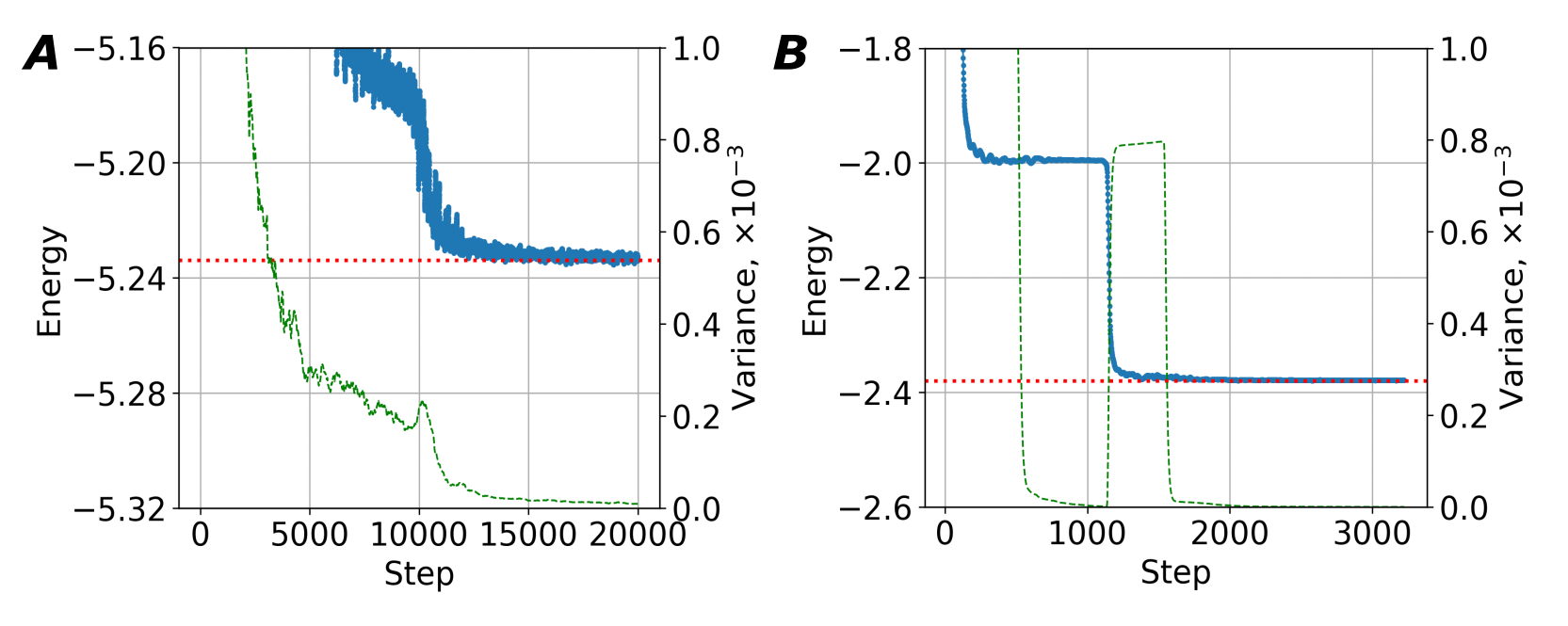}
    \caption{Optimization curves for (A) the Zeeman spin-spiral Hamiltonian, Eq.~\eqref{eq:spin_spiral}, with 3 electrons, showing the moving average of the energy over 5 optimization steps;  (B) the Rashba Hamiltonian, Eq.~\eqref{eq:rashba}, with 5 electrons, showing the moving average of the energy over 20 optimization steps. In both cases, we used two-dimensional period systems with period 6 in each spatial direction and spin update probability $p=0.1$.
     }
    \label{fig:spin-depenedent-simulation}
\end{figure}

In Fig.~\ref{fig:spin-depenedent-simulation}~A, we show the performance of our model for 3 electrons in this Hamiltonian compared to the exact value of energy. The energy expectation of the variational wavefunction reaches the analytically exact ground state energy (to at least 3 significant digits) within 20,000 steps. This demonstrates that our method can accurately capture spatially-dependent spin textures. 

\subsection{Spin-orbit interaction} \label{sec:soc}
The spin-orbit coupling (SOC) is a relativistic correction relevant in some chemical and solid state application. A common example of SOC is the Rashba term:
\bea \label{eq:rashba}
H_{\rm SOC} = p_x \sigma_y - p_y \sigma_x.
\eea

The results of simulation for ${\bf p}^2/2 + H_{\rm SOC}$ are shown in Fig.~\ref{fig:spin-depenedent-simulation}~B. We observe that the analytical ground state energy is reached within 2,000 optimization steps, showing that spin-momentum coupling can also be accurately captured.

\subsection{Antiferromagnetism} \label{sec:afm}

To study antiferromagnetism, we anticipate its emergence in a two-dimensional electron gas with honeycomb potential and Coulomb interaction \cite{ZhangYuanFu2020, ZhangIsobeFu2020DFT,luo2024simulating}: 
\begin{align}
    H & = \sum_i \left( 
-\frac{1}{2}\boldsymbol{\nabla}_i^2 + V(\mathbf{r}_i) 
\right) 
+ \frac{r_{\rm s}}{2} \sum_i \sum_{i \neq j} \frac{1}{|\mathbf{r}_i - \mathbf{r}_j|}, 
\label{eq:system-hamiltonian}
\end{align}
where $V(\mathbf{r}) = -2V_0 \sum_{j=1}^3 \cos(\mathbf{g}_j \cdot \mathbf{r} + \varphi)$ is the potential with reciprocal lattice vectors $\mathbf{g}_j = \frac{4\pi}{\sqrt{3}a_{\rm M}} (\cos \frac{2\pi j}{3}, \sin \frac{2\pi j}{3})$, lattice constant $a_M$, and $\varphi$ controls the shape of the potential \cite{Wu2018}. 
This Hamiltonian is an effective model for two-dimensional $\Gamma$-valley  moir\'e semiconductors in transition metal dichalcogenides \cite{zhang2021tmd}.
We here use $\varphi = \pi$ to realize a honeycomb potential, a moir\'e length $a_{\rm M} = \sqrt{2\pi/\sqrt{3}}$, potential strength $V_0 = 10.0$, and interaction strength $r_{\rm s} = 10$. For these parameters, our results indicate an antiferromagnetic ground state. 

\begin{figure}
    \centering
    \includegraphics[width=\linewidth]{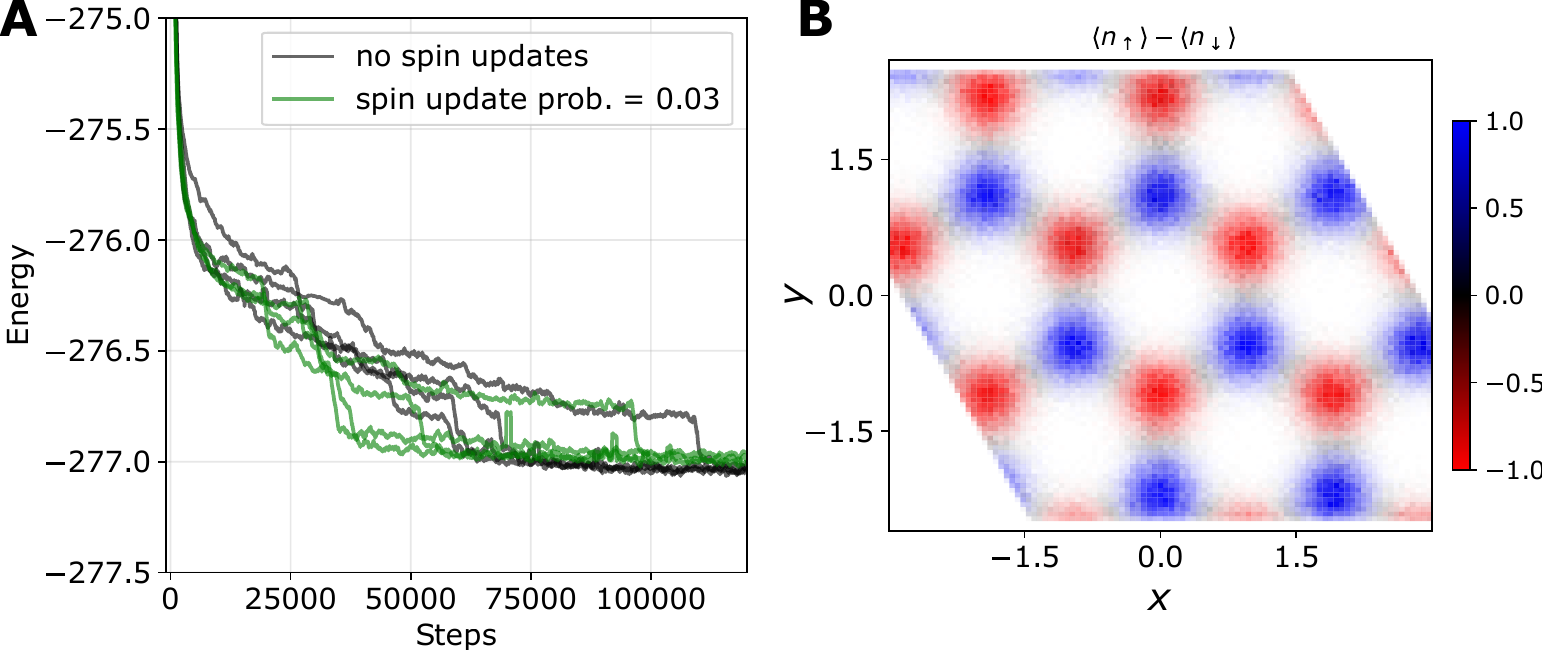}
    \caption{Benchmark of spin updates in the Markov chain Monte Carlo routine at a two-dimensional electron gas with honeycomb potential: (a) Energy as a function of step with (black) no spin updates and (green) spin $s_z$ conserving updates with spin swap probability $p = 0.03$. The figure shows four runs with different random initialization for both Monte-Carlo procedures. (b) Final spin density, where the color indicates spin polarization $\frac{\langle n_\uparrow \rangle -\langle n_\downarrow \rangle}{\langle n_\uparrow \rangle +\langle n_\downarrow \rangle}$ and the saturation total density $\langle n_\uparrow \rangle +\langle n_\downarrow \rangle$. }
    \label{fig:benchmark-AFM-spin-updates}
\end{figure}

Fig.~\ref{fig:benchmark-AFM-spin-updates}(a) shows the energy as a function of optimization step during training of the neural network wavefunction for multiple random initialization seeds, comparing computations without spin updates and with sector-preserving spin swap probability $p_{\rm swap} = 0.03$. On average, including spin swaps reduces the number of training steps required until the antiferromagnetic state is reached (as indicated by the last jump to the approximately constant energy achieved by all curves at large runtime). Both without and with spin updates, the required number of steps until the antiferromagnetic state is achieved varies within an approximate factor of two. The spin density profile confirming the antiferromagnetic order is shown in Fig.~\ref{fig:benchmark-AFM-spin-updates}(b). We verified that all curves achieve an antiferromagnetic spin order equivalent to Fig.~\ref{fig:benchmark-AFM-spin-updates}(b). These results demonstrate that sector-preserving spin updates on average reduce the time until the antiferromagnetic ground state is reached. 


\section{Conclusion and Outlook}
\label{sec:conclusions}

We presented a VMC solver for ground states of spinful fermionic systems. It is based on a neural network ansatz with joint embedding of all DOFs and self-attention to represent the generalized orbitals. Energy estimation is performed with an MCMC that employs separate position and spin updates. Benchmarks on spin-spiral and Rashba SOC models yield accurate ground-state energies and spin textures. We also observed that for an antiferromagnetic ground state in a spin-conserving Hamiltonian, enabling sector-preserving discrete updates accelerated convergence.

This framework naturally extends to include layer, valley, sublattice, or other isospin DOFs which unlocks the simulation of a wide range of condensed matter systems. 
Specifically, the application to 2D materials such as multi-layer graphene \cite{cao2018insulator,cao2018superconductivity,Yankowitz2019superconductivity,zhou2021halfmetal,zhou2022superconductivity-bernal,lu2024fqaheGraphene,lu2025qahe} and transition metal dichalcogenides \cite{xu2020correlated,tang2020hubbard,cai2023fqaheMoTe2,redekop2024fci,Xia2025Jan,Guo2025Jan} is particularly promising because of the variety of correlated phases observed in these materials.

\begin{acknowledgments}
We thank Xiang Li for useful discussions.  This work was supported by a Simons Investigator Award from the Simons Foundation. M.G. acknowledges support from the German Research
Foundation under the Walter Benjamin program (Grant
Agreement No. 526129603). This work made use of resources provided by subMIT at MIT Physics and by the National Science Foundation under Cooperative Agreement PHY-2019786. 
Numerical calculations in this paper build on the recently developed code ``PeriodicWave'', which is publicly available on GitHub \cite{periodicwave_github} and documented on the website {\tt deeppsi.ai}.

\end{acknowledgments}

\bibliography{biblio}

\begin{thebibliography}{71}%
\makeatletter
\providecommand \@ifxundefined [1]{%
 \@ifx{#1\undefined}
}%
\providecommand \@ifnum [1]{%
 \ifnum #1\expandafter \@firstoftwo
 \else \expandafter \@secondoftwo
 \fi
}%
\providecommand \@ifx [1]{%
 \ifx #1\expandafter \@firstoftwo
 \else \expandafter \@secondoftwo
 \fi
}%
\providecommand \natexlab [1]{#1}%
\providecommand \enquote  [1]{``#1''}%
\providecommand \bibnamefont  [1]{#1}%
\providecommand \bibfnamefont [1]{#1}%
\providecommand \citenamefont [1]{#1}%
\providecommand \href@noop [0]{\@secondoftwo}%
\providecommand \href [0]{\begingroup \@sanitize@url \@href}%
\providecommand \@href[1]{\@@startlink{#1}\@@href}%
\providecommand \@@href[1]{\endgroup#1\@@endlink}%
\providecommand \@sanitize@url [0]{\catcode `\\12\catcode `\$12\catcode `\&12\catcode `\#12\catcode `\^12\catcode `\_12\catcode `\%12\relax}%
\providecommand \@@startlink[1]{}%
\providecommand \@@endlink[0]{}%
\providecommand \url  [0]{\begingroup\@sanitize@url \@url }%
\providecommand \@url [1]{\endgroup\@href {#1}{\urlprefix }}%
\providecommand \urlprefix  [0]{URL }%
\providecommand \Eprint [0]{\href }%
\providecommand \doibase [0]{https://doi.org/}%
\providecommand \selectlanguage [0]{\@gobble}%
\providecommand \bibinfo  [0]{\@secondoftwo}%
\providecommand \bibfield  [0]{\@secondoftwo}%
\providecommand \translation [1]{[#1]}%
\providecommand \BibitemOpen [0]{}%
\providecommand \bibitemStop [0]{}%
\providecommand \bibitemNoStop [0]{.\EOS\space}%
\providecommand \EOS [0]{\spacefactor3000\relax}%
\providecommand \BibitemShut  [1]{\csname bibitem#1\endcsname}%
\let\auto@bib@innerbib\@empty
\bibitem [{\citenamefont {Carleo}\ and\ \citenamefont {Troyer}(2017)}]{Carleo2017}%
  \BibitemOpen
  \bibfield  {author} {\bibinfo {author} {\bibfnamefont {G.}~\bibnamefont {Carleo}}\ and\ \bibinfo {author} {\bibfnamefont {M.}~\bibnamefont {Troyer}},\ }\bibfield  {title} {\bibinfo {title} {Solving the quantum many-body problem with artificial neural networks},\ }\href {https://doi.org/10.1126/science.aag2302} {\bibfield  {journal} {\bibinfo  {journal} {Science}\ }\textbf {\bibinfo {volume} {355}},\ \bibinfo {pages} {602} (\bibinfo {year} {2017})},\ \Eprint {https://arxiv.org/abs/https://www.science.org/doi/pdf/10.1126/science.aag2302} {https://www.science.org/doi/pdf/10.1126/science.aag2302} \BibitemShut {NoStop}%
\bibitem [{\citenamefont {Pfau}\ \emph {et~al.}(2020)\citenamefont {Pfau}, \citenamefont {Spencer}, \citenamefont {Matthews},\ and\ \citenamefont {Foulkes}}]{pfau2020ab}%
  \BibitemOpen
  \bibfield  {author} {\bibinfo {author} {\bibfnamefont {D.}~\bibnamefont {Pfau}}, \bibinfo {author} {\bibfnamefont {J.~S.}\ \bibnamefont {Spencer}}, \bibinfo {author} {\bibfnamefont {A.~G.}\ \bibnamefont {Matthews}},\ and\ \bibinfo {author} {\bibfnamefont {W.~M.~C.}\ \bibnamefont {Foulkes}},\ }\bibfield  {title} {\bibinfo {title} {{Ab initio solution of the many-electron Schr{\"o}dinger equation with deep neural networks}},\ }\href@noop {} {\bibfield  {journal} {\bibinfo  {journal} {Physical review research}\ }\textbf {\bibinfo {volume} {2}},\ \bibinfo {pages} {033429} (\bibinfo {year} {2020})}\BibitemShut {NoStop}%
\bibitem [{\citenamefont {Hermann}\ \emph {et~al.}(2020)\citenamefont {Hermann}, \citenamefont {Sch{\"a}tzle},\ and\ \citenamefont {No{\'e}}}]{Hermann2020PauliNet}%
  \BibitemOpen
  \bibfield  {author} {\bibinfo {author} {\bibfnamefont {J.}~\bibnamefont {Hermann}}, \bibinfo {author} {\bibfnamefont {Z.}~\bibnamefont {Sch{\"a}tzle}},\ and\ \bibinfo {author} {\bibfnamefont {F.}~\bibnamefont {No{\'e}}},\ }\bibfield  {title} {\bibinfo {title} {{Deep-neural-network solution of the electronic Schr{\"o}dinger equation}},\ }\href {https://doi.org/10.1038/s41557-020-0544-y} {\bibfield  {journal} {\bibinfo  {journal} {Nature Chemistry}\ }\textbf {\bibinfo {volume} {12}},\ \bibinfo {pages} {891} (\bibinfo {year} {2020})}\BibitemShut {NoStop}%
\bibitem [{\citenamefont {von Glehn}\ \emph {et~al.}(2022)\citenamefont {von Glehn}, \citenamefont {Spencer},\ and\ \citenamefont {Pfau}}]{von2022self}%
  \BibitemOpen
  \bibfield  {author} {\bibinfo {author} {\bibfnamefont {I.}~\bibnamefont {von Glehn}}, \bibinfo {author} {\bibfnamefont {J.~S.}\ \bibnamefont {Spencer}},\ and\ \bibinfo {author} {\bibfnamefont {D.}~\bibnamefont {Pfau}},\ }\bibfield  {title} {\bibinfo {title} {A self-attention ansatz for ab-initio quantum chemistry},\ }\href@noop {} {\bibfield  {journal} {\bibinfo  {journal} {arXiv preprint arXiv:2211.13672}\ } (\bibinfo {year} {2022})}\BibitemShut {NoStop}%
\bibitem [{\citenamefont {Pescia}\ \emph {et~al.}(2022)\citenamefont {Pescia}, \citenamefont {Han}, \citenamefont {Lovato}, \citenamefont {Lu},\ and\ \citenamefont {Carleo}}]{pescia2022neural}%
  \BibitemOpen
  \bibfield  {author} {\bibinfo {author} {\bibfnamefont {G.}~\bibnamefont {Pescia}}, \bibinfo {author} {\bibfnamefont {J.}~\bibnamefont {Han}}, \bibinfo {author} {\bibfnamefont {A.}~\bibnamefont {Lovato}}, \bibinfo {author} {\bibfnamefont {J.}~\bibnamefont {Lu}},\ and\ \bibinfo {author} {\bibfnamefont {G.}~\bibnamefont {Carleo}},\ }\bibfield  {title} {\bibinfo {title} {Neural-network quantum states for periodic systems in continuous space},\ }\href@noop {} {\bibfield  {journal} {\bibinfo  {journal} {Physical Review Research}\ }\textbf {\bibinfo {volume} {4}},\ \bibinfo {pages} {023138} (\bibinfo {year} {2022})}\BibitemShut {NoStop}%
\bibitem [{\citenamefont {Choo}\ \emph {et~al.}(2020)\citenamefont {Choo}, \citenamefont {Mezzacapo},\ and\ \citenamefont {Carleo}}]{choo2020fermionic}%
  \BibitemOpen
  \bibfield  {author} {\bibinfo {author} {\bibfnamefont {K.}~\bibnamefont {Choo}}, \bibinfo {author} {\bibfnamefont {A.}~\bibnamefont {Mezzacapo}},\ and\ \bibinfo {author} {\bibfnamefont {G.}~\bibnamefont {Carleo}},\ }\bibfield  {title} {\bibinfo {title} {Fermionic neural-network states for ab-initio electronic structure},\ }\href@noop {} {\bibfield  {journal} {\bibinfo  {journal} {Nature communications}\ }\textbf {\bibinfo {volume} {11}},\ \bibinfo {pages} {2368} (\bibinfo {year} {2020})}\BibitemShut {NoStop}%
\bibitem [{\citenamefont {Jiang}\ \emph {et~al.}(2025)\citenamefont {Jiang}, \citenamefont {Wen}, \citenamefont {Chen}, \citenamefont {Li}, \citenamefont {Fu}, \citenamefont {Pham}, \citenamefont {Chen}, \citenamefont {He}, \citenamefont {Goddard~III}, \citenamefont {Wang} \emph {et~al.}}]{jiang2025}%
  \BibitemOpen
  \bibfield  {author} {\bibinfo {author} {\bibfnamefont {D.}~\bibnamefont {Jiang}}, \bibinfo {author} {\bibfnamefont {X.}~\bibnamefont {Wen}}, \bibinfo {author} {\bibfnamefont {Y.}~\bibnamefont {Chen}}, \bibinfo {author} {\bibfnamefont {R.}~\bibnamefont {Li}}, \bibinfo {author} {\bibfnamefont {W.}~\bibnamefont {Fu}}, \bibinfo {author} {\bibfnamefont {H.~Q.}\ \bibnamefont {Pham}}, \bibinfo {author} {\bibfnamefont {J.}~\bibnamefont {Chen}}, \bibinfo {author} {\bibfnamefont {D.}~\bibnamefont {He}}, \bibinfo {author} {\bibfnamefont {W.~A.}\ \bibnamefont {Goddard~III}}, \bibinfo {author} {\bibfnamefont {L.}~\bibnamefont {Wang}}, \emph {et~al.},\ }\bibfield  {title} {\bibinfo {title} {{Neural Scaling Laws Surpass Chemical Accuracy for the Many-Electron Schr$\backslash$" odinger Equation}},\ }\href@noop {} {\bibfield  {journal} {\bibinfo  {journal} {arXiv preprint arXiv:2508.02570}\ } (\bibinfo {year} {2025})}\BibitemShut {NoStop}%
\bibitem [{\citenamefont {Foster}\ \emph {et~al.}(2025)\citenamefont {Foster}, \citenamefont {Sch{\"a}tzle}, \citenamefont {Szab{\'o}}, \citenamefont {Cheng}, \citenamefont {K{\"o}hler}, \citenamefont {Cassella}, \citenamefont {Gao}, \citenamefont {Li}, \citenamefont {No{\'e}},\ and\ \citenamefont {Hermann}}]{foster2025ab}%
  \BibitemOpen
  \bibfield  {author} {\bibinfo {author} {\bibfnamefont {A.}~\bibnamefont {Foster}}, \bibinfo {author} {\bibfnamefont {Z.}~\bibnamefont {Sch{\"a}tzle}}, \bibinfo {author} {\bibfnamefont {P.~B.}\ \bibnamefont {Szab{\'o}}}, \bibinfo {author} {\bibfnamefont {L.}~\bibnamefont {Cheng}}, \bibinfo {author} {\bibfnamefont {J.}~\bibnamefont {K{\"o}hler}}, \bibinfo {author} {\bibfnamefont {G.}~\bibnamefont {Cassella}}, \bibinfo {author} {\bibfnamefont {N.}~\bibnamefont {Gao}}, \bibinfo {author} {\bibfnamefont {J.}~\bibnamefont {Li}}, \bibinfo {author} {\bibfnamefont {F.}~\bibnamefont {No{\'e}}},\ and\ \bibinfo {author} {\bibfnamefont {J.}~\bibnamefont {Hermann}},\ }\bibfield  {title} {\bibinfo {title} {An ab initio foundation model of wavefunctions that accurately describes chemical bond breaking},\ }\href@noop {} {\bibfield  {journal} {\bibinfo  {journal} {arXiv preprint arXiv:2506.19960}\ } (\bibinfo {year} {2025})}\BibitemShut {NoStop}%
\bibitem [{\citenamefont {Cassella}\ \emph {et~al.}(2023{\natexlab{a}})\citenamefont {Cassella}, \citenamefont {Sutterud}, \citenamefont {Azadi}, \citenamefont {Drummond}, \citenamefont {Pfau}, \citenamefont {Spencer},\ and\ \citenamefont {Foulkes}}]{cassella2023discovering}%
  \BibitemOpen
  \bibfield  {author} {\bibinfo {author} {\bibfnamefont {G.}~\bibnamefont {Cassella}}, \bibinfo {author} {\bibfnamefont {H.}~\bibnamefont {Sutterud}}, \bibinfo {author} {\bibfnamefont {S.}~\bibnamefont {Azadi}}, \bibinfo {author} {\bibfnamefont {N.~D.}\ \bibnamefont {Drummond}}, \bibinfo {author} {\bibfnamefont {D.}~\bibnamefont {Pfau}}, \bibinfo {author} {\bibfnamefont {J.~S.}\ \bibnamefont {Spencer}},\ and\ \bibinfo {author} {\bibfnamefont {W.~M.~C.}\ \bibnamefont {Foulkes}},\ }\bibfield  {title} {\bibinfo {title} {Discovering quantum phase transitions with fermionic neural networks},\ }\href@noop {} {\bibfield  {journal} {\bibinfo  {journal} {Physical Review Letters}\ }\textbf {\bibinfo {volume} {130}},\ \bibinfo {pages} {036401} (\bibinfo {year} {2023}{\natexlab{a}})}\BibitemShut {NoStop}%
\bibitem [{\citenamefont {Pescia}\ \emph {et~al.}(2024)\citenamefont {Pescia}, \citenamefont {Nys}, \citenamefont {Kim}, \citenamefont {Lovato},\ and\ \citenamefont {Carleo}}]{pescia2024message}%
  \BibitemOpen
  \bibfield  {author} {\bibinfo {author} {\bibfnamefont {G.}~\bibnamefont {Pescia}}, \bibinfo {author} {\bibfnamefont {J.}~\bibnamefont {Nys}}, \bibinfo {author} {\bibfnamefont {J.}~\bibnamefont {Kim}}, \bibinfo {author} {\bibfnamefont {A.}~\bibnamefont {Lovato}},\ and\ \bibinfo {author} {\bibfnamefont {G.}~\bibnamefont {Carleo}},\ }\bibfield  {title} {\bibinfo {title} {Message-passing neural quantum states for the homogeneous electron gas},\ }\href@noop {} {\bibfield  {journal} {\bibinfo  {journal} {Physical Review B}\ }\textbf {\bibinfo {volume} {110}},\ \bibinfo {pages} {035108} (\bibinfo {year} {2024})}\BibitemShut {NoStop}%
\bibitem [{\citenamefont {Li}\ \emph {et~al.}(2025{\natexlab{a}})\citenamefont {Li}, \citenamefont {Ong}, \citenamefont {Geier}, \citenamefont {Lin},\ and\ \citenamefont {Fu}}]{li2025attention}%
  \BibitemOpen
  \bibfield  {author} {\bibinfo {author} {\bibfnamefont {C.-T.}\ \bibnamefont {Li}}, \bibinfo {author} {\bibfnamefont {T.}~\bibnamefont {Ong}}, \bibinfo {author} {\bibfnamefont {M.}~\bibnamefont {Geier}}, \bibinfo {author} {\bibfnamefont {H.}~\bibnamefont {Lin}},\ and\ \bibinfo {author} {\bibfnamefont {L.}~\bibnamefont {Fu}},\ }\bibfield  {title} {\bibinfo {title} {Attention is all you need to solve chiral superconductivity},\ }\href@noop {} {\bibfield  {journal} {\bibinfo  {journal} {arXiv preprint arXiv:2509.03683}\ } (\bibinfo {year} {2025}{\natexlab{a}})}\BibitemShut {NoStop}%
\bibitem [{\citenamefont {Teng}\ \emph {et~al.}(2024)\citenamefont {Teng}, \citenamefont {Dai},\ and\ \citenamefont {Fu}}]{teng2024solving}%
  \BibitemOpen
  \bibfield  {author} {\bibinfo {author} {\bibfnamefont {Y.}~\bibnamefont {Teng}}, \bibinfo {author} {\bibfnamefont {D.~D.}\ \bibnamefont {Dai}},\ and\ \bibinfo {author} {\bibfnamefont {L.}~\bibnamefont {Fu}},\ }\bibfield  {title} {\bibinfo {title} {{Solving and visualizing fractional quantum Hall wavefunctions with neural network}},\ }\href@noop {} {\bibfield  {journal} {\bibinfo  {journal} {arXiv preprint arXiv:2412.00618}\ } (\bibinfo {year} {2024})}\BibitemShut {NoStop}%
\bibitem [{\citenamefont {Qian}\ \emph {et~al.}(2025)\citenamefont {Qian}, \citenamefont {Zhao}, \citenamefont {Zhang}, \citenamefont {Xiang}, \citenamefont {Li},\ and\ \citenamefont {Chen}}]{qian2025describing}%
  \BibitemOpen
  \bibfield  {author} {\bibinfo {author} {\bibfnamefont {Y.}~\bibnamefont {Qian}}, \bibinfo {author} {\bibfnamefont {T.}~\bibnamefont {Zhao}}, \bibinfo {author} {\bibfnamefont {J.}~\bibnamefont {Zhang}}, \bibinfo {author} {\bibfnamefont {T.}~\bibnamefont {Xiang}}, \bibinfo {author} {\bibfnamefont {X.}~\bibnamefont {Li}},\ and\ \bibinfo {author} {\bibfnamefont {J.}~\bibnamefont {Chen}},\ }\bibfield  {title} {\bibinfo {title} {{Describing Landau Level Mixing in Fractional Quantum Hall States with Deep Learning}},\ }\href@noop {} {\bibfield  {journal} {\bibinfo  {journal} {Physical Review Letters}\ }\textbf {\bibinfo {volume} {134}},\ \bibinfo {pages} {176503} (\bibinfo {year} {2025})}\BibitemShut {NoStop}%
\bibitem [{\citenamefont {Nazaryan}\ \emph {et~al.}(2025)\citenamefont {Nazaryan}, \citenamefont {Gaggioli}, \citenamefont {Teng},\ and\ \citenamefont {Fu}}]{nazaryan2025artificial}%
  \BibitemOpen
  \bibfield  {author} {\bibinfo {author} {\bibfnamefont {K.}~\bibnamefont {Nazaryan}}, \bibinfo {author} {\bibfnamefont {F.}~\bibnamefont {Gaggioli}}, \bibinfo {author} {\bibfnamefont {Y.}~\bibnamefont {Teng}},\ and\ \bibinfo {author} {\bibfnamefont {L.}~\bibnamefont {Fu}},\ }\bibfield  {title} {\bibinfo {title} {{Artificial Intelligence for Quantum Matter: Finding a Needle in a Haystack}},\ }\href@noop {} {\bibfield  {journal} {\bibinfo  {journal} {arXiv preprint arXiv:2507.13322}\ } (\bibinfo {year} {2025})}\BibitemShut {NoStop}%
\bibitem [{\citenamefont {Luo}\ \emph {et~al.}(2025)\citenamefont {Luo}, \citenamefont {Zaklama},\ and\ \citenamefont {Fu}}]{luo2025solving}%
  \BibitemOpen
  \bibfield  {author} {\bibinfo {author} {\bibfnamefont {D.}~\bibnamefont {Luo}}, \bibinfo {author} {\bibfnamefont {T.}~\bibnamefont {Zaklama}},\ and\ \bibinfo {author} {\bibfnamefont {L.}~\bibnamefont {Fu}},\ }\bibfield  {title} {\bibinfo {title} {{Solving fractional electron states in twisted MoTe $ \_2 $ with deep neural network}},\ }\href@noop {} {\bibfield  {journal} {\bibinfo  {journal} {arXiv preprint arXiv:2503.13585}\ } (\bibinfo {year} {2025})}\BibitemShut {NoStop}%
\bibitem [{\citenamefont {Li}\ \emph {et~al.}(2025{\natexlab{b}})\citenamefont {Li}, \citenamefont {Chen}, \citenamefont {Li}, \citenamefont {Chen}, \citenamefont {Wu}, \citenamefont {Chen},\ and\ \citenamefont {Ren}}]{li2025deep}%
  \BibitemOpen
  \bibfield  {author} {\bibinfo {author} {\bibfnamefont {X.}~\bibnamefont {Li}}, \bibinfo {author} {\bibfnamefont {Y.}~\bibnamefont {Chen}}, \bibinfo {author} {\bibfnamefont {B.}~\bibnamefont {Li}}, \bibinfo {author} {\bibfnamefont {H.}~\bibnamefont {Chen}}, \bibinfo {author} {\bibfnamefont {F.}~\bibnamefont {Wu}}, \bibinfo {author} {\bibfnamefont {J.}~\bibnamefont {Chen}},\ and\ \bibinfo {author} {\bibfnamefont {W.}~\bibnamefont {Ren}},\ }\bibfield  {title} {\bibinfo {title} {{Deep Learning Sheds Light on Integer and Fractional Topological Insulators}},\ }\href@noop {} {\bibfield  {journal} {\bibinfo  {journal} {arXiv preprint arXiv:2503.11756}\ } (\bibinfo {year} {2025}{\natexlab{b}})}\BibitemShut {NoStop}%
\bibitem [{\citenamefont {Carleo}\ \emph {et~al.}(2019)\citenamefont {Carleo}, \citenamefont {Choo}, \citenamefont {Hofmann}, \citenamefont {Smith}, \citenamefont {Westerhout}, \citenamefont {Alet}, \citenamefont {Davis}, \citenamefont {Efthymiou}, \citenamefont {Glasser}, \citenamefont {Lin} \emph {et~al.}}]{carleo2019netket}%
  \BibitemOpen
  \bibfield  {author} {\bibinfo {author} {\bibfnamefont {G.}~\bibnamefont {Carleo}}, \bibinfo {author} {\bibfnamefont {K.}~\bibnamefont {Choo}}, \bibinfo {author} {\bibfnamefont {D.}~\bibnamefont {Hofmann}}, \bibinfo {author} {\bibfnamefont {J.~E.}\ \bibnamefont {Smith}}, \bibinfo {author} {\bibfnamefont {T.}~\bibnamefont {Westerhout}}, \bibinfo {author} {\bibfnamefont {F.}~\bibnamefont {Alet}}, \bibinfo {author} {\bibfnamefont {E.~J.}\ \bibnamefont {Davis}}, \bibinfo {author} {\bibfnamefont {S.}~\bibnamefont {Efthymiou}}, \bibinfo {author} {\bibfnamefont {I.}~\bibnamefont {Glasser}}, \bibinfo {author} {\bibfnamefont {S.-H.}\ \bibnamefont {Lin}}, \emph {et~al.},\ }\bibfield  {title} {\bibinfo {title} {{NetKet: A machine learning toolkit for many-body quantum systems}},\ }\href@noop {} {\bibfield  {journal} {\bibinfo  {journal} {SoftwareX}\ }\textbf {\bibinfo {volume} {10}},\ \bibinfo {pages} {100311} (\bibinfo {year} {2019})}\BibitemShut {NoStop}%
\bibitem [{\citenamefont {Gu}\ \emph {et~al.}(2025{\natexlab{a}})\citenamefont {Gu}, \citenamefont {Li}, \citenamefont {Lin}, \citenamefont {Zhan}, \citenamefont {Li}, \citenamefont {Huang}, \citenamefont {He}, \citenamefont {Wu}, \citenamefont {Xiang}, \citenamefont {Qin} \emph {et~al.}}]{gu2025solving}%
  \BibitemOpen
  \bibfield  {author} {\bibinfo {author} {\bibfnamefont {Y.}~\bibnamefont {Gu}}, \bibinfo {author} {\bibfnamefont {W.}~\bibnamefont {Li}}, \bibinfo {author} {\bibfnamefont {H.}~\bibnamefont {Lin}}, \bibinfo {author} {\bibfnamefont {B.}~\bibnamefont {Zhan}}, \bibinfo {author} {\bibfnamefont {R.}~\bibnamefont {Li}}, \bibinfo {author} {\bibfnamefont {Y.}~\bibnamefont {Huang}}, \bibinfo {author} {\bibfnamefont {D.}~\bibnamefont {He}}, \bibinfo {author} {\bibfnamefont {Y.}~\bibnamefont {Wu}}, \bibinfo {author} {\bibfnamefont {T.}~\bibnamefont {Xiang}}, \bibinfo {author} {\bibfnamefont {M.}~\bibnamefont {Qin}}, \emph {et~al.},\ }\bibfield  {title} {\bibinfo {title} {{Solving the Hubbard model with Neural Quantum States}},\ }\href@noop {} {\bibfield  {journal} {\bibinfo  {journal} {arXiv preprint arXiv:2507.02644}\ } (\bibinfo {year} {2025}{\natexlab{a}})}\BibitemShut {NoStop}%
\bibitem [{\citenamefont {Luo}\ and\ \citenamefont {Clark}(2019)}]{luo2019backflow}%
  \BibitemOpen
  \bibfield  {author} {\bibinfo {author} {\bibfnamefont {D.}~\bibnamefont {Luo}}\ and\ \bibinfo {author} {\bibfnamefont {B.~K.}\ \bibnamefont {Clark}},\ }\bibfield  {title} {\bibinfo {title} {Backflow transformations via neural networks for quantum many-body wave functions},\ }\href@noop {} {\bibfield  {journal} {\bibinfo  {journal} {Physical review letters}\ }\textbf {\bibinfo {volume} {122}},\ \bibinfo {pages} {226401} (\bibinfo {year} {2019})}\BibitemShut {NoStop}%
\bibitem [{\citenamefont {Robledo~Moreno}\ \emph {et~al.}(2022)\citenamefont {Robledo~Moreno}, \citenamefont {Carleo}, \citenamefont {Georges},\ and\ \citenamefont {Stokes}}]{robledo2022fermionic}%
  \BibitemOpen
  \bibfield  {author} {\bibinfo {author} {\bibfnamefont {J.}~\bibnamefont {Robledo~Moreno}}, \bibinfo {author} {\bibfnamefont {G.}~\bibnamefont {Carleo}}, \bibinfo {author} {\bibfnamefont {A.}~\bibnamefont {Georges}},\ and\ \bibinfo {author} {\bibfnamefont {J.}~\bibnamefont {Stokes}},\ }\bibfield  {title} {\bibinfo {title} {Fermionic wave functions from neural-network constrained hidden states},\ }\href@noop {} {\bibfield  {journal} {\bibinfo  {journal} {Proceedings of the National Academy of Sciences}\ }\textbf {\bibinfo {volume} {119}},\ \bibinfo {pages} {e2122059119} (\bibinfo {year} {2022})}\BibitemShut {NoStop}%
\bibitem [{\citenamefont {Roth}\ \emph {et~al.}(2023)\citenamefont {Roth}, \citenamefont {Szab\'o},\ and\ \citenamefont {MacDonald}}]{PhysRevB.108.054410}%
  \BibitemOpen
  \bibfield  {author} {\bibinfo {author} {\bibfnamefont {C.}~\bibnamefont {Roth}}, \bibinfo {author} {\bibfnamefont {A.}~\bibnamefont {Szab\'o}},\ and\ \bibinfo {author} {\bibfnamefont {A.~H.}\ \bibnamefont {MacDonald}},\ }\bibfield  {title} {\bibinfo {title} {{High-accuracy variational Monte Carlo for frustrated magnets with deep neural networks}},\ }\href {https://doi.org/10.1103/PhysRevB.108.054410} {\bibfield  {journal} {\bibinfo  {journal} {Phys. Rev. B}\ }\textbf {\bibinfo {volume} {108}},\ \bibinfo {pages} {054410} (\bibinfo {year} {2023})}\BibitemShut {NoStop}%
\bibitem [{\citenamefont {Astrakhantsev}\ \emph {et~al.}(2021)\citenamefont {Astrakhantsev}, \citenamefont {Westerhout}, \citenamefont {Tiwari}, \citenamefont {Choo}, \citenamefont {Chen}, \citenamefont {Fischer}, \citenamefont {Carleo},\ and\ \citenamefont {Neupert}}]{astrakhantsev2021broken}%
  \BibitemOpen
  \bibfield  {author} {\bibinfo {author} {\bibfnamefont {N.}~\bibnamefont {Astrakhantsev}}, \bibinfo {author} {\bibfnamefont {T.}~\bibnamefont {Westerhout}}, \bibinfo {author} {\bibfnamefont {A.}~\bibnamefont {Tiwari}}, \bibinfo {author} {\bibfnamefont {K.}~\bibnamefont {Choo}}, \bibinfo {author} {\bibfnamefont {A.}~\bibnamefont {Chen}}, \bibinfo {author} {\bibfnamefont {M.~H.}\ \bibnamefont {Fischer}}, \bibinfo {author} {\bibfnamefont {G.}~\bibnamefont {Carleo}},\ and\ \bibinfo {author} {\bibfnamefont {T.}~\bibnamefont {Neupert}},\ }\bibfield  {title} {\bibinfo {title} {{Broken-symmetry ground states of the Heisenberg model on the pyrochlore lattice}},\ }\href@noop {} {\bibfield  {journal} {\bibinfo  {journal} {Physical Review X}\ }\textbf {\bibinfo {volume} {11}},\ \bibinfo {pages} {041021} (\bibinfo {year} {2021})}\BibitemShut {NoStop}%
\bibitem [{\citenamefont {Choo}\ \emph {et~al.}(2019)\citenamefont {Choo}, \citenamefont {Neupert},\ and\ \citenamefont {Carleo}}]{choo2019two}%
  \BibitemOpen
  \bibfield  {author} {\bibinfo {author} {\bibfnamefont {K.}~\bibnamefont {Choo}}, \bibinfo {author} {\bibfnamefont {T.}~\bibnamefont {Neupert}},\ and\ \bibinfo {author} {\bibfnamefont {G.}~\bibnamefont {Carleo}},\ }\bibfield  {title} {\bibinfo {title} {{Two-dimensional frustrated J 1-J 2 model studied with neural network quantum states}},\ }\href@noop {} {\bibfield  {journal} {\bibinfo  {journal} {Physical Review B}\ }\textbf {\bibinfo {volume} {100}},\ \bibinfo {pages} {125124} (\bibinfo {year} {2019})}\BibitemShut {NoStop}%
\bibitem [{\citenamefont {Adams}\ \emph {et~al.}(2021)\citenamefont {Adams}, \citenamefont {Carleo}, \citenamefont {Lovato},\ and\ \citenamefont {Rocco}}]{adams2021variational}%
  \BibitemOpen
  \bibfield  {author} {\bibinfo {author} {\bibfnamefont {C.}~\bibnamefont {Adams}}, \bibinfo {author} {\bibfnamefont {G.}~\bibnamefont {Carleo}}, \bibinfo {author} {\bibfnamefont {A.}~\bibnamefont {Lovato}},\ and\ \bibinfo {author} {\bibfnamefont {N.}~\bibnamefont {Rocco}},\ }\bibfield  {title} {\bibinfo {title} {{Variational Monte Carlo calculations of $A\leq4$ nuclei with an artificial neural-network correlator ansatz}},\ }\href@noop {} {\bibfield  {journal} {\bibinfo  {journal} {Physical Review Letters}\ }\textbf {\bibinfo {volume} {127}},\ \bibinfo {pages} {022502} (\bibinfo {year} {2021})}\BibitemShut {NoStop}%
\bibitem [{\citenamefont {Zhan}\ \emph {et~al.}(2025)\citenamefont {Zhan}, \citenamefont {Wheeler}, \citenamefont {Ertekin}, \citenamefont {Adams},\ and\ \citenamefont {Wagner}}]{zhan2025}%
  \BibitemOpen
  \bibfield  {author} {\bibinfo {author} {\bibfnamefont {N.}~\bibnamefont {Zhan}}, \bibinfo {author} {\bibfnamefont {W.~A.}\ \bibnamefont {Wheeler}}, \bibinfo {author} {\bibfnamefont {E.}~\bibnamefont {Ertekin}}, \bibinfo {author} {\bibfnamefont {R.~P.}\ \bibnamefont {Adams}},\ and\ \bibinfo {author} {\bibfnamefont {L.~K.}\ \bibnamefont {Wagner}},\ }\bibfield  {title} {\bibinfo {title} {Expressivity of determinantal anzatzes for neural network wave functions},\ }\href@noop {} {\bibfield  {journal} {\bibinfo  {journal} {arXiv preprint arXiv:2506.00155}\ } (\bibinfo {year} {2025})}\BibitemShut {NoStop}%
\bibitem [{\citenamefont {Baltz}\ \emph {et~al.}(2018)\citenamefont {Baltz}, \citenamefont {Manchon}, \citenamefont {Tsoi}, \citenamefont {Moriyama}, \citenamefont {Ono},\ and\ \citenamefont {Tserkovnyak}}]{spintronics_baltz}%
  \BibitemOpen
  \bibfield  {author} {\bibinfo {author} {\bibfnamefont {V.}~\bibnamefont {Baltz}}, \bibinfo {author} {\bibfnamefont {A.}~\bibnamefont {Manchon}}, \bibinfo {author} {\bibfnamefont {M.}~\bibnamefont {Tsoi}}, \bibinfo {author} {\bibfnamefont {T.}~\bibnamefont {Moriyama}}, \bibinfo {author} {\bibfnamefont {T.}~\bibnamefont {Ono}},\ and\ \bibinfo {author} {\bibfnamefont {Y.}~\bibnamefont {Tserkovnyak}},\ }\bibfield  {title} {\bibinfo {title} {Antiferromagnetic spintronics},\ }\href {https://doi.org/10.1103/RevModPhys.90.015005} {\bibfield  {journal} {\bibinfo  {journal} {Rev. Mod. Phys.}\ }\textbf {\bibinfo {volume} {90}},\ \bibinfo {pages} {015005} (\bibinfo {year} {2018})}\BibitemShut {NoStop}%
\bibitem [{\citenamefont {Spaldin}\ and\ \citenamefont {Fiebig}(2005)}]{multiferroics_spaldin}%
  \BibitemOpen
  \bibfield  {author} {\bibinfo {author} {\bibfnamefont {N.~A.}\ \bibnamefont {Spaldin}}\ and\ \bibinfo {author} {\bibfnamefont {M.}~\bibnamefont {Fiebig}},\ }\bibfield  {title} {\bibinfo {title} {{The Renaissance of Magnetoelectric Multiferroics}},\ }\href {https://doi.org/10.1126/science.1113357} {\bibfield  {journal} {\bibinfo  {journal} {Science}\ }\textbf {\bibinfo {volume} {309}},\ \bibinfo {pages} {391} (\bibinfo {year} {2005})},\ \Eprint {https://arxiv.org/abs/https://www.science.org/doi/pdf/10.1126/science.1113357} {https://www.science.org/doi/pdf/10.1126/science.1113357} \BibitemShut {NoStop}%
\bibitem [{\citenamefont {Nagaosa}\ and\ \citenamefont {Tokura}(2013)}]{skyrmions_nagaosa}%
  \BibitemOpen
  \bibfield  {author} {\bibinfo {author} {\bibfnamefont {N.}~\bibnamefont {Nagaosa}}\ and\ \bibinfo {author} {\bibfnamefont {Y.}~\bibnamefont {Tokura}},\ }\bibfield  {title} {\bibinfo {title} {Topological properties and dynamics of magnetic skyrmions},\ }\href {https://doi.org/10.1038/nnano.2013.243} {\bibfield  {journal} {\bibinfo  {journal} {Nature Nanotechnology}\ }\textbf {\bibinfo {volume} {8}},\ \bibinfo {pages} {899} (\bibinfo {year} {2013})}\BibitemShut {NoStop}%
\bibitem [{\citenamefont {Moore}(2010)}]{moore2010birth}%
  \BibitemOpen
  \bibfield  {author} {\bibinfo {author} {\bibfnamefont {J.~E.}\ \bibnamefont {Moore}},\ }\bibfield  {title} {\bibinfo {title} {The birth of topological insulators},\ }\href@noop {} {\bibfield  {journal} {\bibinfo  {journal} {Nature}\ }\textbf {\bibinfo {volume} {464}},\ \bibinfo {pages} {194} (\bibinfo {year} {2010})}\BibitemShut {NoStop}%
\bibitem [{\citenamefont {Fu}\ and\ \citenamefont {Kane}(2007)}]{fu2007topological}%
  \BibitemOpen
  \bibfield  {author} {\bibinfo {author} {\bibfnamefont {L.}~\bibnamefont {Fu}}\ and\ \bibinfo {author} {\bibfnamefont {C.~L.}\ \bibnamefont {Kane}},\ }\bibfield  {title} {\bibinfo {title} {Topological insulators with inversion symmetry},\ }\href@noop {} {\bibfield  {journal} {\bibinfo  {journal} {Physical Review B—Condensed Matter and Materials Physics}\ }\textbf {\bibinfo {volume} {76}},\ \bibinfo {pages} {045302} (\bibinfo {year} {2007})}\BibitemShut {NoStop}%
\bibitem [{\citenamefont {Hasan}\ and\ \citenamefont {Kane}(2010)}]{hasan2010colloquium}%
  \BibitemOpen
  \bibfield  {author} {\bibinfo {author} {\bibfnamefont {M.~Z.}\ \bibnamefont {Hasan}}\ and\ \bibinfo {author} {\bibfnamefont {C.~L.}\ \bibnamefont {Kane}},\ }\bibfield  {title} {\bibinfo {title} {Colloquium: topological insulators},\ }\href@noop {} {\bibfield  {journal} {\bibinfo  {journal} {Reviews of modern physics}\ }\textbf {\bibinfo {volume} {82}},\ \bibinfo {pages} {3045} (\bibinfo {year} {2010})}\BibitemShut {NoStop}%
\bibitem [{\citenamefont {Armitage}\ \emph {et~al.}(2018)\citenamefont {Armitage}, \citenamefont {Mele},\ and\ \citenamefont {Vishwanath}}]{armitage2018weyl}%
  \BibitemOpen
  \bibfield  {author} {\bibinfo {author} {\bibfnamefont {N.~P.}\ \bibnamefont {Armitage}}, \bibinfo {author} {\bibfnamefont {E.~J.}\ \bibnamefont {Mele}},\ and\ \bibinfo {author} {\bibfnamefont {A.}~\bibnamefont {Vishwanath}},\ }\bibfield  {title} {\bibinfo {title} {{Weyl and Dirac semimetals in three-dimensional solids}},\ }\href@noop {} {\bibfield  {journal} {\bibinfo  {journal} {Reviews of Modern Physics}\ }\textbf {\bibinfo {volume} {90}},\ \bibinfo {pages} {015001} (\bibinfo {year} {2018})}\BibitemShut {NoStop}%
\bibitem [{\citenamefont {Melton}\ \emph {et~al.}(2016{\natexlab{a}})\citenamefont {Melton}, \citenamefont {Bennett},\ and\ \citenamefont {Mitas}}]{Melton2016QMCspin}%
  \BibitemOpen
  \bibfield  {author} {\bibinfo {author} {\bibfnamefont {C.~A.}\ \bibnamefont {Melton}}, \bibinfo {author} {\bibfnamefont {M.~C.}\ \bibnamefont {Bennett}},\ and\ \bibinfo {author} {\bibfnamefont {L.}~\bibnamefont {Mitas}},\ }\bibfield  {title} {\bibinfo {title} {{Quantum Monte Carlo with variable spins}},\ }\href {https://doi.org/10.1063/1.4954726} {\bibfield  {journal} {\bibinfo  {journal} {The Journal of Chemical Physics}\ }\textbf {\bibinfo {volume} {144}},\ \bibinfo {pages} {244113} (\bibinfo {year} {2016}{\natexlab{a}})}\BibitemShut {NoStop}%
\bibitem [{\citenamefont {Melton}\ \emph {et~al.}(2016{\natexlab{b}})\citenamefont {Melton}, \citenamefont {Zhu}, \citenamefont {Guo}, \citenamefont {Ambrosetti}, \citenamefont {Pederiva},\ and\ \citenamefont {Mitas}}]{Melton2016spinorbit}%
  \BibitemOpen
  \bibfield  {author} {\bibinfo {author} {\bibfnamefont {C.~A.}\ \bibnamefont {Melton}}, \bibinfo {author} {\bibfnamefont {M.}~\bibnamefont {Zhu}}, \bibinfo {author} {\bibfnamefont {S.}~\bibnamefont {Guo}}, \bibinfo {author} {\bibfnamefont {A.}~\bibnamefont {Ambrosetti}}, \bibinfo {author} {\bibfnamefont {F.}~\bibnamefont {Pederiva}},\ and\ \bibinfo {author} {\bibfnamefont {L.}~\bibnamefont {Mitas}},\ }\bibfield  {title} {\bibinfo {title} {{Spin-orbit interactions in electronic structure quantum Monte Carlo methods}},\ }\href {https://doi.org/10.1103/PhysRevA.93.042502} {\bibfield  {journal} {\bibinfo  {journal} {Phys. Rev. A}\ }\textbf {\bibinfo {volume} {93}},\ \bibinfo {pages} {042502} (\bibinfo {year} {2016}{\natexlab{b}})}\BibitemShut {NoStop}%
\bibitem [{\citenamefont {Gerard}\ \emph {et~al.}(2024)\citenamefont {Gerard}, \citenamefont {Scherbela}, \citenamefont {Sutterud}, \citenamefont {Foulkes},\ and\ \citenamefont {Grohs}}]{gerard2024solids}%
  \BibitemOpen
  \bibfield  {author} {\bibinfo {author} {\bibfnamefont {L.}~\bibnamefont {Gerard}}, \bibinfo {author} {\bibfnamefont {M.}~\bibnamefont {Scherbela}}, \bibinfo {author} {\bibfnamefont {H.}~\bibnamefont {Sutterud}}, \bibinfo {author} {\bibfnamefont {M.}~\bibnamefont {Foulkes}},\ and\ \bibinfo {author} {\bibfnamefont {P.}~\bibnamefont {Grohs}},\ }\href {https://arxiv.org/abs/2405.07599} {\bibinfo {title} {{Transferable Neural Wavefunctions for Solids}}} (\bibinfo {year} {2024}),\ \Eprint {https://arxiv.org/abs/2405.07599} {arXiv:2405.07599 [physics.comp-ph]} \BibitemShut {NoStop}%
\bibitem [{\citenamefont {Vaswani}\ \emph {et~al.}(2017)\citenamefont {Vaswani}, \citenamefont {Shazeer}, \citenamefont {Parmar}, \citenamefont {Uszkoreit}, \citenamefont {Jones}, \citenamefont {Gomez}, \citenamefont {Kaiser},\ and\ \citenamefont {Polosukhin}}]{vaswani2017attention}%
  \BibitemOpen
  \bibfield  {author} {\bibinfo {author} {\bibfnamefont {A.}~\bibnamefont {Vaswani}}, \bibinfo {author} {\bibfnamefont {N.}~\bibnamefont {Shazeer}}, \bibinfo {author} {\bibfnamefont {N.}~\bibnamefont {Parmar}}, \bibinfo {author} {\bibfnamefont {J.}~\bibnamefont {Uszkoreit}}, \bibinfo {author} {\bibfnamefont {L.}~\bibnamefont {Jones}}, \bibinfo {author} {\bibfnamefont {A.~N.}\ \bibnamefont {Gomez}}, \bibinfo {author} {\bibfnamefont {{\L}.}~\bibnamefont {Kaiser}},\ and\ \bibinfo {author} {\bibfnamefont {I.}~\bibnamefont {Polosukhin}},\ }\bibfield  {title} {\bibinfo {title} {Attention is all you need},\ }\href@noop {} {\bibfield  {journal} {\bibinfo  {journal} {Advances in neural information processing systems}\ }\textbf {\bibinfo {volume} {30}} (\bibinfo {year} {2017})}\BibitemShut {NoStop}%
\bibitem [{\citenamefont {Geier}\ \emph {et~al.}(2025)\citenamefont {Geier}, \citenamefont {Nazaryan}, \citenamefont {Zaklama},\ and\ \citenamefont {Fu}}]{geier2025attention}%
  \BibitemOpen
  \bibfield  {author} {\bibinfo {author} {\bibfnamefont {M.}~\bibnamefont {Geier}}, \bibinfo {author} {\bibfnamefont {K.}~\bibnamefont {Nazaryan}}, \bibinfo {author} {\bibfnamefont {T.}~\bibnamefont {Zaklama}},\ and\ \bibinfo {author} {\bibfnamefont {L.}~\bibnamefont {Fu}},\ }\bibfield  {title} {\bibinfo {title} {Is attention all you need to solve the correlated electron problem?},\ }\href@noop {} {\bibfield  {journal} {\bibinfo  {journal} {arXiv preprint arXiv:2502.05383}\ } (\bibinfo {year} {2025})}\BibitemShut {NoStop}%
\bibitem [{\citenamefont {Li}\ \emph {et~al.}(2025{\natexlab{c}})\citenamefont {Li}, \citenamefont {Ong}, \citenamefont {Geier}, \citenamefont {Lin},\ and\ \citenamefont {Fu}}]{li2025attentionneedsolvechiral}%
  \BibitemOpen
  \bibfield  {author} {\bibinfo {author} {\bibfnamefont {C.-T.}\ \bibnamefont {Li}}, \bibinfo {author} {\bibfnamefont {T.}~\bibnamefont {Ong}}, \bibinfo {author} {\bibfnamefont {M.}~\bibnamefont {Geier}}, \bibinfo {author} {\bibfnamefont {H.}~\bibnamefont {Lin}},\ and\ \bibinfo {author} {\bibfnamefont {L.}~\bibnamefont {Fu}},\ }\href {https://arxiv.org/abs/2509.03683} {\bibinfo {title} {Attention is all you need to solve chiral superconductivity}} (\bibinfo {year} {2025}{\natexlab{c}}),\ \Eprint {https://arxiv.org/abs/2509.03683} {arXiv:2509.03683 [cond-mat.supr-con]} \BibitemShut {NoStop}%
\bibitem [{\citenamefont {Amari}(1998)}]{Amari1998NGD}%
  \BibitemOpen
  \bibfield  {author} {\bibinfo {author} {\bibfnamefont {S.-i.}\ \bibnamefont {Amari}},\ }\bibfield  {title} {\bibinfo {title} {{Natural Gradient Works Efficiently in Learning}},\ }\href {https://doi.org/10.1162/089976698300017746} {\bibfield  {journal} {\bibinfo  {journal} {Neural Computation}\ }\textbf {\bibinfo {volume} {10}},\ \bibinfo {pages} {251} (\bibinfo {year} {1998})}\BibitemShut {NoStop}%
\bibitem [{\citenamefont {Sorella}(1998)}]{Sorella1998}%
  \BibitemOpen
  \bibfield  {author} {\bibinfo {author} {\bibfnamefont {S.}~\bibnamefont {Sorella}},\ }\bibfield  {title} {\bibinfo {title} {{Green Function Monte Carlo with Stochastic Reconfiguration}},\ }\href {https://doi.org/10.1103/PhysRevLett.80.4558} {\bibfield  {journal} {\bibinfo  {journal} {Phys. Rev. Lett.}\ }\textbf {\bibinfo {volume} {80}},\ \bibinfo {pages} {4558} (\bibinfo {year} {1998})}\BibitemShut {NoStop}%
\bibitem [{\citenamefont {Stokes}\ \emph {et~al.}(2020)\citenamefont {Stokes}, \citenamefont {Izaac}, \citenamefont {Killoran},\ and\ \citenamefont {Carleo}}]{Stokes2020quantumnatural}%
  \BibitemOpen
  \bibfield  {author} {\bibinfo {author} {\bibfnamefont {J.}~\bibnamefont {Stokes}}, \bibinfo {author} {\bibfnamefont {J.}~\bibnamefont {Izaac}}, \bibinfo {author} {\bibfnamefont {N.}~\bibnamefont {Killoran}},\ and\ \bibinfo {author} {\bibfnamefont {G.}~\bibnamefont {Carleo}},\ }\bibfield  {title} {\bibinfo {title} {{Quantum {N}atural {G}radient}},\ }\href {https://doi.org/10.22331/q-2020-05-25-269} {\bibfield  {journal} {\bibinfo  {journal} {{Quantum}}\ }\textbf {\bibinfo {volume} {4}},\ \bibinfo {pages} {269} (\bibinfo {year} {2020})}\BibitemShut {NoStop}%
\bibitem [{\citenamefont {Martens}\ and\ \citenamefont {Grosse}(2015)}]{Martens2015KFAC}%
  \BibitemOpen
  \bibfield  {author} {\bibinfo {author} {\bibfnamefont {J.}~\bibnamefont {Martens}}\ and\ \bibinfo {author} {\bibfnamefont {R.}~\bibnamefont {Grosse}},\ }\bibfield  {title} {\bibinfo {title} {{Optimizing neural networks with Kronecker-factored approximate curvature}},\ }in\ \href@noop {} {\emph {\bibinfo {booktitle} {Proceedings of the 32nd International Conference on International Conference on Machine Learning - Volume 37}}},\ \bibinfo {series and number} {ICML'15}\ (\bibinfo  {publisher} {JMLR.org},\ \bibinfo {year} {2015})\ p.\ \bibinfo {pages} {2408–2417}\BibitemShut {NoStop}%
\bibitem [{\citenamefont {Chen}\ and\ \citenamefont {Heyl}(2024)}]{Chen2024minSR}%
  \BibitemOpen
  \bibfield  {author} {\bibinfo {author} {\bibfnamefont {A.}~\bibnamefont {Chen}}\ and\ \bibinfo {author} {\bibfnamefont {M.}~\bibnamefont {Heyl}},\ }\bibfield  {title} {\bibinfo {title} {Empowering deep neural quantum states through efficient optimization},\ }\href {https://doi.org/10.1038/s41567-024-02566-1} {\bibfield  {journal} {\bibinfo  {journal} {Nature Physics}\ }\textbf {\bibinfo {volume} {20}},\ \bibinfo {pages} {1476} (\bibinfo {year} {2024})}\BibitemShut {NoStop}%
\bibitem [{\citenamefont {Goldshlager}\ \emph {et~al.}(2024)\citenamefont {Goldshlager}, \citenamefont {Abrahamsen},\ and\ \citenamefont {Lin}}]{Goldshlager2024SPRING}%
  \BibitemOpen
  \bibfield  {author} {\bibinfo {author} {\bibfnamefont {G.}~\bibnamefont {Goldshlager}}, \bibinfo {author} {\bibfnamefont {N.}~\bibnamefont {Abrahamsen}},\ and\ \bibinfo {author} {\bibfnamefont {L.}~\bibnamefont {Lin}},\ }\bibfield  {title} {\bibinfo {title} {{A Kaczmarz-inspired approach to accelerate the optimization of neural network wavefunctions}},\ }\href {https://doi.org/https://doi.org/10.1016/j.jcp.2024.113351} {\bibfield  {journal} {\bibinfo  {journal} {Journal of Computational Physics}\ }\textbf {\bibinfo {volume} {516}},\ \bibinfo {pages} {113351} (\bibinfo {year} {2024})}\BibitemShut {NoStop}%
\bibitem [{\citenamefont {Gu}\ \emph {et~al.}(2025{\natexlab{b}})\citenamefont {Gu}, \citenamefont {Li}, \citenamefont {Lin}, \citenamefont {Zhan}, \citenamefont {Li}, \citenamefont {Huang}, \citenamefont {He}, \citenamefont {Wu}, \citenamefont {Xiang}, \citenamefont {Qin}, \citenamefont {Wang},\ and\ \citenamefont {Lv}}]{gu2025solvinghubbardmodelneural}%
  \BibitemOpen
  \bibfield  {author} {\bibinfo {author} {\bibfnamefont {Y.}~\bibnamefont {Gu}}, \bibinfo {author} {\bibfnamefont {W.}~\bibnamefont {Li}}, \bibinfo {author} {\bibfnamefont {H.}~\bibnamefont {Lin}}, \bibinfo {author} {\bibfnamefont {B.}~\bibnamefont {Zhan}}, \bibinfo {author} {\bibfnamefont {R.}~\bibnamefont {Li}}, \bibinfo {author} {\bibfnamefont {Y.}~\bibnamefont {Huang}}, \bibinfo {author} {\bibfnamefont {D.}~\bibnamefont {He}}, \bibinfo {author} {\bibfnamefont {Y.}~\bibnamefont {Wu}}, \bibinfo {author} {\bibfnamefont {T.}~\bibnamefont {Xiang}}, \bibinfo {author} {\bibfnamefont {M.}~\bibnamefont {Qin}}, \bibinfo {author} {\bibfnamefont {L.}~\bibnamefont {Wang}},\ and\ \bibinfo {author} {\bibfnamefont {D.}~\bibnamefont {Lv}},\ }\href {https://arxiv.org/abs/2507.02644} {\bibinfo {title} {{Solving the Hubbard model with Neural Quantum States}}} (\bibinfo {year} {2025}{\natexlab{b}}),\ \Eprint {https://arxiv.org/abs/2507.02644} {arXiv:2507.02644 [cond-mat.str-el]} \BibitemShut {NoStop}%
\bibitem [{\citenamefont {Kwon}\ \emph {et~al.}(1993)\citenamefont {Kwon}, \citenamefont {Ceperley},\ and\ \citenamefont {Martin}}]{kwon1993backflow}%
  \BibitemOpen
  \bibfield  {author} {\bibinfo {author} {\bibfnamefont {Y.}~\bibnamefont {Kwon}}, \bibinfo {author} {\bibfnamefont {D.~M.}\ \bibnamefont {Ceperley}},\ and\ \bibinfo {author} {\bibfnamefont {R.~M.}\ \bibnamefont {Martin}},\ }\bibfield  {title} {\bibinfo {title} {Effects of three-body and backflow correlations in the two-dimensional electron gas},\ }\href {https://doi.org/10.1103/PhysRevB.48.12037} {\bibfield  {journal} {\bibinfo  {journal} {Phys. Rev. B}\ }\textbf {\bibinfo {volume} {48}},\ \bibinfo {pages} {12037} (\bibinfo {year} {1993})}\BibitemShut {NoStop}%
\bibitem [{\citenamefont {Feynman}\ and\ \citenamefont {Cohen}(1956)}]{Feynman1956}%
  \BibitemOpen
  \bibfield  {author} {\bibinfo {author} {\bibfnamefont {R.~P.}\ \bibnamefont {Feynman}}\ and\ \bibinfo {author} {\bibfnamefont {M.}~\bibnamefont {Cohen}},\ }\bibfield  {title} {\bibinfo {title} {{Energy Spectrum of the Excitations in Liquid Helium}},\ }\href {https://doi.org/10.1103/PhysRev.102.1189} {\bibfield  {journal} {\bibinfo  {journal} {Phys. Rev.}\ }\textbf {\bibinfo {volume} {102}},\ \bibinfo {pages} {1189} (\bibinfo {year} {1956})}\BibitemShut {NoStop}%
\bibitem [{\citenamefont {Yun}\ \emph {et~al.}(2020)\citenamefont {Yun}, \citenamefont {Bhojanapalli}, \citenamefont {Rawat}, \citenamefont {Reddi},\ and\ \citenamefont {Kumar}}]{Yun2020Are}%
  \BibitemOpen
  \bibfield  {author} {\bibinfo {author} {\bibfnamefont {C.}~\bibnamefont {Yun}}, \bibinfo {author} {\bibfnamefont {S.}~\bibnamefont {Bhojanapalli}}, \bibinfo {author} {\bibfnamefont {A.~S.}\ \bibnamefont {Rawat}}, \bibinfo {author} {\bibfnamefont {S.}~\bibnamefont {Reddi}},\ and\ \bibinfo {author} {\bibfnamefont {S.}~\bibnamefont {Kumar}},\ }\bibfield  {title} {\bibinfo {title} {{Are Transformers universal approximators of sequence-to-sequence functions?}},\ }in\ \href {https://openreview.net/forum?id=ByxRM0Ntvr} {\emph {\bibinfo {booktitle} {International Conference on Learning Representations}}}\ (\bibinfo {year} {2020})\BibitemShut {NoStop}%
\bibitem [{\citenamefont {Foulkes}\ \emph {et~al.}(2001)\citenamefont {Foulkes}, \citenamefont {Mitas}, \citenamefont {Needs},\ and\ \citenamefont {Rajagopal}}]{Foulkes_review}%
  \BibitemOpen
  \bibfield  {author} {\bibinfo {author} {\bibfnamefont {W.~M.~C.}\ \bibnamefont {Foulkes}}, \bibinfo {author} {\bibfnamefont {L.}~\bibnamefont {Mitas}}, \bibinfo {author} {\bibfnamefont {R.~J.}\ \bibnamefont {Needs}},\ and\ \bibinfo {author} {\bibfnamefont {G.}~\bibnamefont {Rajagopal}},\ }\bibfield  {title} {\bibinfo {title} {{Quantum Monte Carlo simulations of solids}},\ }\href {https://doi.org/10.1103/RevModPhys.73.33} {\bibfield  {journal} {\bibinfo  {journal} {Rev. Mod. Phys.}\ }\textbf {\bibinfo {volume} {73}},\ \bibinfo {pages} {33} (\bibinfo {year} {2001})}\BibitemShut {NoStop}%
\bibitem [{\citenamefont {Li}\ \emph {et~al.}(2023)\citenamefont {Li}, \citenamefont {Ye}, \citenamefont {Jiang}, \citenamefont {Wen}, \citenamefont {Wang}, \citenamefont {Li}, \citenamefont {Li}, \citenamefont {He}, \citenamefont {Chen}, \citenamefont {Ren} \emph {et~al.}}]{li2023forward}%
  \BibitemOpen
  \bibfield  {author} {\bibinfo {author} {\bibfnamefont {R.}~\bibnamefont {Li}}, \bibinfo {author} {\bibfnamefont {H.}~\bibnamefont {Ye}}, \bibinfo {author} {\bibfnamefont {D.}~\bibnamefont {Jiang}}, \bibinfo {author} {\bibfnamefont {X.}~\bibnamefont {Wen}}, \bibinfo {author} {\bibfnamefont {C.}~\bibnamefont {Wang}}, \bibinfo {author} {\bibfnamefont {Z.}~\bibnamefont {Li}}, \bibinfo {author} {\bibfnamefont {X.}~\bibnamefont {Li}}, \bibinfo {author} {\bibfnamefont {D.}~\bibnamefont {He}}, \bibinfo {author} {\bibfnamefont {J.}~\bibnamefont {Chen}}, \bibinfo {author} {\bibfnamefont {W.}~\bibnamefont {Ren}}, \emph {et~al.},\ }\bibfield  {title} {\bibinfo {title} {{Forward laplacian: A new computational framework for neural network-based variational Monte Carlo}},\ }\href@noop {} {\bibfield  {journal} {\bibinfo  {journal} {arXiv preprint arXiv:2307.08214}\ } (\bibinfo {year} {2023})}\BibitemShut {NoStop}%
\bibitem [{\citenamefont {Onishi}\ \emph {et~al.}(2025)\citenamefont {Onishi}, \citenamefont {Paul},\ and\ \citenamefont {Fu}}]{onishi2025emergent}%
  \BibitemOpen
  \bibfield  {author} {\bibinfo {author} {\bibfnamefont {Y.}~\bibnamefont {Onishi}}, \bibinfo {author} {\bibfnamefont {N.}~\bibnamefont {Paul}},\ and\ \bibinfo {author} {\bibfnamefont {L.}~\bibnamefont {Fu}},\ }\bibfield  {title} {\bibinfo {title} {Emergent gravity and gravitational lensing in quantum materials},\ }\href@noop {} {\bibfield  {journal} {\bibinfo  {journal} {arXiv preprint arXiv:2506.04335}\ } (\bibinfo {year} {2025})}\BibitemShut {NoStop}%
\bibitem [{\citenamefont {Zhang}\ \emph {et~al.}(2020{\natexlab{a}})\citenamefont {Zhang}, \citenamefont {Yuan},\ and\ \citenamefont {Fu}}]{ZhangYuanFu2020}%
  \BibitemOpen
  \bibfield  {author} {\bibinfo {author} {\bibfnamefont {Y.}~\bibnamefont {Zhang}}, \bibinfo {author} {\bibfnamefont {N.~F.~Q.}\ \bibnamefont {Yuan}},\ and\ \bibinfo {author} {\bibfnamefont {L.}~\bibnamefont {Fu}},\ }\bibfield  {title} {\bibinfo {title} {Moir\'e quantum chemistry: Charge transfer in transition metal dichalcogenide superlattices},\ }\href {https://doi.org/10.1103/PhysRevB.102.201115} {\bibfield  {journal} {\bibinfo  {journal} {Phys. Rev. B}\ }\textbf {\bibinfo {volume} {102}},\ \bibinfo {pages} {201115} (\bibinfo {year} {2020}{\natexlab{a}})}\BibitemShut {NoStop}%
\bibitem [{\citenamefont {Zhang}\ \emph {et~al.}(2020{\natexlab{b}})\citenamefont {Zhang}, \citenamefont {Isobe},\ and\ \citenamefont {Fu}}]{ZhangIsobeFu2020DFT}%
  \BibitemOpen
  \bibfield  {author} {\bibinfo {author} {\bibfnamefont {Y.}~\bibnamefont {Zhang}}, \bibinfo {author} {\bibfnamefont {H.}~\bibnamefont {Isobe}},\ and\ \bibinfo {author} {\bibfnamefont {L.}~\bibnamefont {Fu}},\ }\href {https://arxiv.org/abs/2005.04238} {\bibinfo {title} {Density functional approach to correlated moire states: itinerant magnetism}} (\bibinfo {year} {2020}{\natexlab{b}}),\ \Eprint {https://arxiv.org/abs/2005.04238} {arXiv:2005.04238 [cond-mat.str-el]} \BibitemShut {NoStop}%
\bibitem [{\citenamefont {Luo}\ \emph {et~al.}(2024)\citenamefont {Luo}, \citenamefont {Dai},\ and\ \citenamefont {Fu}}]{luo2024simulating}%
  \BibitemOpen
  \bibfield  {author} {\bibinfo {author} {\bibfnamefont {D.}~\bibnamefont {Luo}}, \bibinfo {author} {\bibfnamefont {D.~D.}\ \bibnamefont {Dai}},\ and\ \bibinfo {author} {\bibfnamefont {L.}~\bibnamefont {Fu}},\ }\bibfield  {title} {\bibinfo {title} {Simulating moir$\backslash$'e quantum matter with neural network},\ }\href@noop {} {\bibfield  {journal} {\bibinfo  {journal} {arXiv preprint arXiv:2406.17645}\ } (\bibinfo {year} {2024})}\BibitemShut {NoStop}%
\bibitem [{\citenamefont {Wu}\ \emph {et~al.}(2018)\citenamefont {Wu}, \citenamefont {Lovorn}, \citenamefont {Tutuc},\ and\ \citenamefont {MacDonald}}]{Wu2018}%
  \BibitemOpen
  \bibfield  {author} {\bibinfo {author} {\bibfnamefont {F.}~\bibnamefont {Wu}}, \bibinfo {author} {\bibfnamefont {T.}~\bibnamefont {Lovorn}}, \bibinfo {author} {\bibfnamefont {E.}~\bibnamefont {Tutuc}},\ and\ \bibinfo {author} {\bibfnamefont {A.~H.}\ \bibnamefont {MacDonald}},\ }\bibfield  {title} {\bibinfo {title} {{Hubbard Model Physics in Transition Metal Dichalcogenide Moir\'e Bands}},\ }\href {https://doi.org/10.1103/PhysRevLett.121.026402} {\bibfield  {journal} {\bibinfo  {journal} {Phys. Rev. Lett.}\ }\textbf {\bibinfo {volume} {121}},\ \bibinfo {pages} {026402} (\bibinfo {year} {2018})}\BibitemShut {NoStop}%
\bibitem [{\citenamefont {Zhang}\ \emph {et~al.}(2021)\citenamefont {Zhang}, \citenamefont {Liu},\ and\ \citenamefont {Fu}}]{zhang2021tmd}%
  \BibitemOpen
  \bibfield  {author} {\bibinfo {author} {\bibfnamefont {Y.}~\bibnamefont {Zhang}}, \bibinfo {author} {\bibfnamefont {T.}~\bibnamefont {Liu}},\ and\ \bibinfo {author} {\bibfnamefont {L.}~\bibnamefont {Fu}},\ }\bibfield  {title} {\bibinfo {title} {Electronic structures, charge transfer, and charge order in twisted transition metal dichalcogenide bilayers},\ }\href {https://doi.org/10.1103/PhysRevB.103.155142} {\bibfield  {journal} {\bibinfo  {journal} {Phys. Rev. B}\ }\textbf {\bibinfo {volume} {103}},\ \bibinfo {pages} {155142} (\bibinfo {year} {2021})}\BibitemShut {NoStop}%
\bibitem [{\citenamefont {Cao}\ \emph {et~al.}(2018{\natexlab{a}})\citenamefont {Cao}, \citenamefont {Fatemi}, \citenamefont {Demir}, \citenamefont {Fang}, \citenamefont {Tomarken}, \citenamefont {Luo}, \citenamefont {Sanchez-Yamagishi}, \citenamefont {Watanabe}, \citenamefont {Taniguchi}, \citenamefont {Kaxiras}, \citenamefont {Ashoori},\ and\ \citenamefont {Jarillo-Herrero}}]{cao2018insulator}%
  \BibitemOpen
  \bibfield  {author} {\bibinfo {author} {\bibfnamefont {Y.}~\bibnamefont {Cao}}, \bibinfo {author} {\bibfnamefont {V.}~\bibnamefont {Fatemi}}, \bibinfo {author} {\bibfnamefont {A.}~\bibnamefont {Demir}}, \bibinfo {author} {\bibfnamefont {S.}~\bibnamefont {Fang}}, \bibinfo {author} {\bibfnamefont {S.~L.}\ \bibnamefont {Tomarken}}, \bibinfo {author} {\bibfnamefont {J.~Y.}\ \bibnamefont {Luo}}, \bibinfo {author} {\bibfnamefont {J.~D.}\ \bibnamefont {Sanchez-Yamagishi}}, \bibinfo {author} {\bibfnamefont {K.}~\bibnamefont {Watanabe}}, \bibinfo {author} {\bibfnamefont {T.}~\bibnamefont {Taniguchi}}, \bibinfo {author} {\bibfnamefont {E.}~\bibnamefont {Kaxiras}}, \bibinfo {author} {\bibfnamefont {R.~C.}\ \bibnamefont {Ashoori}},\ and\ \bibinfo {author} {\bibfnamefont {P.}~\bibnamefont {Jarillo-Herrero}},\ }\bibfield  {title} {\bibinfo {title} {Correlated insulator behaviour at half-filling in magic-angle graphene superlattices},\ }\href {https://doi.org/10.1038/nature26154} {\bibfield  {journal} {\bibinfo  {journal}
  {Nature}\ }\textbf {\bibinfo {volume} {556}},\ \bibinfo {pages} {80} (\bibinfo {year} {2018}{\natexlab{a}})}\BibitemShut {NoStop}%
\bibitem [{\citenamefont {Cao}\ \emph {et~al.}(2018{\natexlab{b}})\citenamefont {Cao}, \citenamefont {Fatemi}, \citenamefont {Fang}, \citenamefont {Watanabe}, \citenamefont {Taniguchi}, \citenamefont {Kaxiras},\ and\ \citenamefont {Jarillo-Herrero}}]{cao2018superconductivity}%
  \BibitemOpen
  \bibfield  {author} {\bibinfo {author} {\bibfnamefont {Y.}~\bibnamefont {Cao}}, \bibinfo {author} {\bibfnamefont {V.}~\bibnamefont {Fatemi}}, \bibinfo {author} {\bibfnamefont {S.}~\bibnamefont {Fang}}, \bibinfo {author} {\bibfnamefont {K.}~\bibnamefont {Watanabe}}, \bibinfo {author} {\bibfnamefont {T.}~\bibnamefont {Taniguchi}}, \bibinfo {author} {\bibfnamefont {E.}~\bibnamefont {Kaxiras}},\ and\ \bibinfo {author} {\bibfnamefont {P.}~\bibnamefont {Jarillo-Herrero}},\ }\bibfield  {title} {\bibinfo {title} {Unconventional superconductivity in magic-angle graphene superlattices},\ }\href {https://doi.org/10.1038/nature26160} {\bibfield  {journal} {\bibinfo  {journal} {Nature}\ }\textbf {\bibinfo {volume} {556}},\ \bibinfo {pages} {43} (\bibinfo {year} {2018}{\natexlab{b}})}\BibitemShut {NoStop}%
\bibitem [{\citenamefont {Yankowitz}\ \emph {et~al.}(2019)\citenamefont {Yankowitz}, \citenamefont {Chen}, \citenamefont {Polshyn}, \citenamefont {Zhang}, \citenamefont {Watanabe}, \citenamefont {Taniguchi}, \citenamefont {Graf}, \citenamefont {Young},\ and\ \citenamefont {Dean}}]{Yankowitz2019superconductivity}%
  \BibitemOpen
  \bibfield  {author} {\bibinfo {author} {\bibfnamefont {M.}~\bibnamefont {Yankowitz}}, \bibinfo {author} {\bibfnamefont {S.}~\bibnamefont {Chen}}, \bibinfo {author} {\bibfnamefont {H.}~\bibnamefont {Polshyn}}, \bibinfo {author} {\bibfnamefont {Y.}~\bibnamefont {Zhang}}, \bibinfo {author} {\bibfnamefont {K.}~\bibnamefont {Watanabe}}, \bibinfo {author} {\bibfnamefont {T.}~\bibnamefont {Taniguchi}}, \bibinfo {author} {\bibfnamefont {D.}~\bibnamefont {Graf}}, \bibinfo {author} {\bibfnamefont {A.~F.}\ \bibnamefont {Young}},\ and\ \bibinfo {author} {\bibfnamefont {C.~R.}\ \bibnamefont {Dean}},\ }\bibfield  {title} {\bibinfo {title} {Tuning superconductivity in twisted bilayer graphene},\ }\href {https://doi.org/10.1126/science.aav1910} {\bibfield  {journal} {\bibinfo  {journal} {Science}\ }\textbf {\bibinfo {volume} {363}},\ \bibinfo {pages} {1059} (\bibinfo {year} {2019})},\ \Eprint {https://arxiv.org/abs/https://www.science.org/doi/pdf/10.1126/science.aav1910}
  {https://www.science.org/doi/pdf/10.1126/science.aav1910} \BibitemShut {NoStop}%
\bibitem [{\citenamefont {Zhou}\ \emph {et~al.}(2021)\citenamefont {Zhou}, \citenamefont {Xie}, \citenamefont {Ghazaryan}, \citenamefont {Holder}, \citenamefont {Ehrets}, \citenamefont {Spanton}, \citenamefont {Taniguchi}, \citenamefont {Watanabe}, \citenamefont {Berg}, \citenamefont {Serbyn},\ and\ \citenamefont {Young}}]{zhou2021halfmetal}%
  \BibitemOpen
  \bibfield  {author} {\bibinfo {author} {\bibfnamefont {H.}~\bibnamefont {Zhou}}, \bibinfo {author} {\bibfnamefont {T.}~\bibnamefont {Xie}}, \bibinfo {author} {\bibfnamefont {A.}~\bibnamefont {Ghazaryan}}, \bibinfo {author} {\bibfnamefont {T.}~\bibnamefont {Holder}}, \bibinfo {author} {\bibfnamefont {J.~R.}\ \bibnamefont {Ehrets}}, \bibinfo {author} {\bibfnamefont {E.~M.}\ \bibnamefont {Spanton}}, \bibinfo {author} {\bibfnamefont {T.}~\bibnamefont {Taniguchi}}, \bibinfo {author} {\bibfnamefont {K.}~\bibnamefont {Watanabe}}, \bibinfo {author} {\bibfnamefont {E.}~\bibnamefont {Berg}}, \bibinfo {author} {\bibfnamefont {M.}~\bibnamefont {Serbyn}},\ and\ \bibinfo {author} {\bibfnamefont {A.~F.}\ \bibnamefont {Young}},\ }\bibfield  {title} {\bibinfo {title} {Half- and quarter-metals in rhombohedral trilayer graphene},\ }\href {https://doi.org/10.1038/s41586-021-03938-w} {\bibfield  {journal} {\bibinfo  {journal} {Nature}\ }\textbf {\bibinfo {volume} {598}},\ \bibinfo {pages} {429} (\bibinfo {year}
  {2021})}\BibitemShut {NoStop}%
\bibitem [{\citenamefont {Zhou}\ \emph {et~al.}(2022)\citenamefont {Zhou}, \citenamefont {Holleis}, \citenamefont {Saito}, \citenamefont {Cohen}, \citenamefont {Huynh}, \citenamefont {Patterson}, \citenamefont {Yang}, \citenamefont {Taniguchi}, \citenamefont {Watanabe},\ and\ \citenamefont {Young}}]{zhou2022superconductivity-bernal}%
  \BibitemOpen
  \bibfield  {author} {\bibinfo {author} {\bibfnamefont {H.}~\bibnamefont {Zhou}}, \bibinfo {author} {\bibfnamefont {L.}~\bibnamefont {Holleis}}, \bibinfo {author} {\bibfnamefont {Y.}~\bibnamefont {Saito}}, \bibinfo {author} {\bibfnamefont {L.}~\bibnamefont {Cohen}}, \bibinfo {author} {\bibfnamefont {W.}~\bibnamefont {Huynh}}, \bibinfo {author} {\bibfnamefont {C.~L.}\ \bibnamefont {Patterson}}, \bibinfo {author} {\bibfnamefont {F.}~\bibnamefont {Yang}}, \bibinfo {author} {\bibfnamefont {T.}~\bibnamefont {Taniguchi}}, \bibinfo {author} {\bibfnamefont {K.}~\bibnamefont {Watanabe}},\ and\ \bibinfo {author} {\bibfnamefont {A.~F.}\ \bibnamefont {Young}},\ }\bibfield  {title} {\bibinfo {title} {{Isospin magnetism and spin-polarized superconductivity in Bernal bilayer graphene}},\ }\href {https://doi.org/10.1126/science.abm8386} {\bibfield  {journal} {\bibinfo  {journal} {Science}\ }\textbf {\bibinfo {volume} {375}},\ \bibinfo {pages} {774} (\bibinfo {year} {2022})},\ \Eprint
  {https://arxiv.org/abs/https://www.science.org/doi/pdf/10.1126/science.abm8386} {https://www.science.org/doi/pdf/10.1126/science.abm8386} \BibitemShut {NoStop}%
\bibitem [{\citenamefont {Lu}\ \emph {et~al.}(2024)\citenamefont {Lu}, \citenamefont {Han}, \citenamefont {Yao}, \citenamefont {Reddy}, \citenamefont {Yang}, \citenamefont {Seo}, \citenamefont {Watanabe}, \citenamefont {Taniguchi}, \citenamefont {Fu},\ and\ \citenamefont {Ju}}]{lu2024fqaheGraphene}%
  \BibitemOpen
  \bibfield  {author} {\bibinfo {author} {\bibfnamefont {Z.}~\bibnamefont {Lu}}, \bibinfo {author} {\bibfnamefont {T.}~\bibnamefont {Han}}, \bibinfo {author} {\bibfnamefont {Y.}~\bibnamefont {Yao}}, \bibinfo {author} {\bibfnamefont {A.~P.}\ \bibnamefont {Reddy}}, \bibinfo {author} {\bibfnamefont {J.}~\bibnamefont {Yang}}, \bibinfo {author} {\bibfnamefont {J.}~\bibnamefont {Seo}}, \bibinfo {author} {\bibfnamefont {K.}~\bibnamefont {Watanabe}}, \bibinfo {author} {\bibfnamefont {T.}~\bibnamefont {Taniguchi}}, \bibinfo {author} {\bibfnamefont {L.}~\bibnamefont {Fu}},\ and\ \bibinfo {author} {\bibfnamefont {L.}~\bibnamefont {Ju}},\ }\bibfield  {title} {\bibinfo {title} {Fractional quantum anomalous hall effect in multilayer graphene},\ }\href {https://doi.org/10.1038/s41586-023-07010-7} {\bibfield  {journal} {\bibinfo  {journal} {Nature}\ }\textbf {\bibinfo {volume} {626}},\ \bibinfo {pages} {759} (\bibinfo {year} {2024})}\BibitemShut {NoStop}%
\bibitem [{\citenamefont {Lu}\ \emph {et~al.}(2025)\citenamefont {Lu}, \citenamefont {Han}, \citenamefont {Yao}, \citenamefont {Hadjri}, \citenamefont {Yang}, \citenamefont {Seo}, \citenamefont {Shi}, \citenamefont {Ye}, \citenamefont {Watanabe}, \citenamefont {Taniguchi},\ and\ \citenamefont {Ju}}]{lu2025qahe}%
  \BibitemOpen
  \bibfield  {author} {\bibinfo {author} {\bibfnamefont {Z.}~\bibnamefont {Lu}}, \bibinfo {author} {\bibfnamefont {T.}~\bibnamefont {Han}}, \bibinfo {author} {\bibfnamefont {Y.}~\bibnamefont {Yao}}, \bibinfo {author} {\bibfnamefont {Z.}~\bibnamefont {Hadjri}}, \bibinfo {author} {\bibfnamefont {J.}~\bibnamefont {Yang}}, \bibinfo {author} {\bibfnamefont {J.}~\bibnamefont {Seo}}, \bibinfo {author} {\bibfnamefont {L.}~\bibnamefont {Shi}}, \bibinfo {author} {\bibfnamefont {S.}~\bibnamefont {Ye}}, \bibinfo {author} {\bibfnamefont {K.}~\bibnamefont {Watanabe}}, \bibinfo {author} {\bibfnamefont {T.}~\bibnamefont {Taniguchi}},\ and\ \bibinfo {author} {\bibfnamefont {L.}~\bibnamefont {Ju}},\ }\bibfield  {title} {\bibinfo {title} {{Extended quantum anomalous Hall states in graphene/hBN Moir{\'e} superlattices}},\ }\href {https://doi.org/10.1038/s41586-024-08470-1} {\bibfield  {journal} {\bibinfo  {journal} {Nature}\ }\textbf {\bibinfo {volume} {637}},\ \bibinfo {pages} {1090} (\bibinfo {year} {2025})}\BibitemShut
  {NoStop}%
\bibitem [{\citenamefont {Xu}\ \emph {et~al.}(2020)\citenamefont {Xu}, \citenamefont {Liu}, \citenamefont {Rhodes}, \citenamefont {Watanabe}, \citenamefont {Taniguchi}, \citenamefont {Hone}, \citenamefont {Elser}, \citenamefont {Mak},\ and\ \citenamefont {Shan}}]{xu2020correlated}%
  \BibitemOpen
  \bibfield  {author} {\bibinfo {author} {\bibfnamefont {Y.}~\bibnamefont {Xu}}, \bibinfo {author} {\bibfnamefont {S.}~\bibnamefont {Liu}}, \bibinfo {author} {\bibfnamefont {D.~A.}\ \bibnamefont {Rhodes}}, \bibinfo {author} {\bibfnamefont {K.}~\bibnamefont {Watanabe}}, \bibinfo {author} {\bibfnamefont {T.}~\bibnamefont {Taniguchi}}, \bibinfo {author} {\bibfnamefont {J.}~\bibnamefont {Hone}}, \bibinfo {author} {\bibfnamefont {V.}~\bibnamefont {Elser}}, \bibinfo {author} {\bibfnamefont {K.~F.}\ \bibnamefont {Mak}},\ and\ \bibinfo {author} {\bibfnamefont {J.}~\bibnamefont {Shan}},\ }\bibfield  {title} {\bibinfo {title} {{Correlated insulating states at fractional fillings of Moir{\'e} superlattices}},\ }\href {https://doi.org/10.1038/s41586-020-2868-6} {\bibfield  {journal} {\bibinfo  {journal} {Nature}\ }\textbf {\bibinfo {volume} {587}},\ \bibinfo {pages} {214} (\bibinfo {year} {2020})}\BibitemShut {NoStop}%
\bibitem [{\citenamefont {Tang}\ \emph {et~al.}(2020)\citenamefont {Tang}, \citenamefont {Li}, \citenamefont {Li}, \citenamefont {Xu}, \citenamefont {Liu}, \citenamefont {Barmak}, \citenamefont {Watanabe}, \citenamefont {Taniguchi}, \citenamefont {MacDonald}, \citenamefont {Shan},\ and\ \citenamefont {Mak}}]{tang2020hubbard}%
  \BibitemOpen
  \bibfield  {author} {\bibinfo {author} {\bibfnamefont {Y.}~\bibnamefont {Tang}}, \bibinfo {author} {\bibfnamefont {L.}~\bibnamefont {Li}}, \bibinfo {author} {\bibfnamefont {T.}~\bibnamefont {Li}}, \bibinfo {author} {\bibfnamefont {Y.}~\bibnamefont {Xu}}, \bibinfo {author} {\bibfnamefont {S.}~\bibnamefont {Liu}}, \bibinfo {author} {\bibfnamefont {K.}~\bibnamefont {Barmak}}, \bibinfo {author} {\bibfnamefont {K.}~\bibnamefont {Watanabe}}, \bibinfo {author} {\bibfnamefont {T.}~\bibnamefont {Taniguchi}}, \bibinfo {author} {\bibfnamefont {A.~H.}\ \bibnamefont {MacDonald}}, \bibinfo {author} {\bibfnamefont {J.}~\bibnamefont {Shan}},\ and\ \bibinfo {author} {\bibfnamefont {K.~F.}\ \bibnamefont {Mak}},\ }\bibfield  {title} {\bibinfo {title} {{Simulation of Hubbard model physics in WSe$_2$/WS$_2$ Moir{\'e} superlattices}},\ }\href {https://doi.org/10.1038/s41586-020-2085-3} {\bibfield  {journal} {\bibinfo  {journal} {Nature}\ }\textbf {\bibinfo {volume} {579}},\ \bibinfo {pages} {353} (\bibinfo {year}
  {2020})}\BibitemShut {NoStop}%
\bibitem [{\citenamefont {Cai}\ \emph {et~al.}(2023)\citenamefont {Cai}, \citenamefont {Anderson}, \citenamefont {Wang}, \citenamefont {Zhang}, \citenamefont {Liu}, \citenamefont {Holtzmann}, \citenamefont {Zhang}, \citenamefont {Fan}, \citenamefont {Taniguchi}, \citenamefont {Watanabe}, \citenamefont {Ran}, \citenamefont {Cao}, \citenamefont {Fu}, \citenamefont {Xiao}, \citenamefont {Yao},\ and\ \citenamefont {Xu}}]{cai2023fqaheMoTe2}%
  \BibitemOpen
  \bibfield  {author} {\bibinfo {author} {\bibfnamefont {J.}~\bibnamefont {Cai}}, \bibinfo {author} {\bibfnamefont {E.}~\bibnamefont {Anderson}}, \bibinfo {author} {\bibfnamefont {C.}~\bibnamefont {Wang}}, \bibinfo {author} {\bibfnamefont {X.}~\bibnamefont {Zhang}}, \bibinfo {author} {\bibfnamefont {X.}~\bibnamefont {Liu}}, \bibinfo {author} {\bibfnamefont {W.}~\bibnamefont {Holtzmann}}, \bibinfo {author} {\bibfnamefont {Y.}~\bibnamefont {Zhang}}, \bibinfo {author} {\bibfnamefont {F.}~\bibnamefont {Fan}}, \bibinfo {author} {\bibfnamefont {T.}~\bibnamefont {Taniguchi}}, \bibinfo {author} {\bibfnamefont {K.}~\bibnamefont {Watanabe}}, \bibinfo {author} {\bibfnamefont {Y.}~\bibnamefont {Ran}}, \bibinfo {author} {\bibfnamefont {T.}~\bibnamefont {Cao}}, \bibinfo {author} {\bibfnamefont {L.}~\bibnamefont {Fu}}, \bibinfo {author} {\bibfnamefont {D.}~\bibnamefont {Xiao}}, \bibinfo {author} {\bibfnamefont {W.}~\bibnamefont {Yao}},\ and\ \bibinfo {author} {\bibfnamefont {X.}~\bibnamefont {Xu}},\ }\bibfield  {title}
  {\bibinfo {title} {{Signatures of fractional quantum anomalous Hall states in twisted MoTe$_2$}},\ }\href {https://doi.org/10.1038/s41586-023-06289-w} {\bibfield  {journal} {\bibinfo  {journal} {Nature}\ }\textbf {\bibinfo {volume} {622}},\ \bibinfo {pages} {63} (\bibinfo {year} {2023})}\BibitemShut {NoStop}%
\bibitem [{\citenamefont {Redekop}\ \emph {et~al.}(2024)\citenamefont {Redekop}, \citenamefont {Zhang}, \citenamefont {Park}, \citenamefont {Cai}, \citenamefont {Anderson}, \citenamefont {Sheekey}, \citenamefont {Arp}, \citenamefont {Babikyan}, \citenamefont {Salters}, \citenamefont {Watanabe}, \citenamefont {Taniguchi}, \citenamefont {Huber}, \citenamefont {Xu},\ and\ \citenamefont {Young}}]{redekop2024fci}%
  \BibitemOpen
  \bibfield  {author} {\bibinfo {author} {\bibfnamefont {E.}~\bibnamefont {Redekop}}, \bibinfo {author} {\bibfnamefont {C.}~\bibnamefont {Zhang}}, \bibinfo {author} {\bibfnamefont {H.}~\bibnamefont {Park}}, \bibinfo {author} {\bibfnamefont {J.}~\bibnamefont {Cai}}, \bibinfo {author} {\bibfnamefont {E.}~\bibnamefont {Anderson}}, \bibinfo {author} {\bibfnamefont {O.}~\bibnamefont {Sheekey}}, \bibinfo {author} {\bibfnamefont {T.}~\bibnamefont {Arp}}, \bibinfo {author} {\bibfnamefont {G.}~\bibnamefont {Babikyan}}, \bibinfo {author} {\bibfnamefont {S.}~\bibnamefont {Salters}}, \bibinfo {author} {\bibfnamefont {K.}~\bibnamefont {Watanabe}}, \bibinfo {author} {\bibfnamefont {T.}~\bibnamefont {Taniguchi}}, \bibinfo {author} {\bibfnamefont {M.~E.}\ \bibnamefont {Huber}}, \bibinfo {author} {\bibfnamefont {X.}~\bibnamefont {Xu}},\ and\ \bibinfo {author} {\bibfnamefont {A.~F.}\ \bibnamefont {Young}},\ }\bibfield  {title} {\bibinfo {title} {{Direct magnetic imaging of fractional Chern insulators in twisted MoTe$_2$}},\
  }\href {https://doi.org/10.1038/s41586-024-08153-x} {\bibfield  {journal} {\bibinfo  {journal} {Nature}\ }\textbf {\bibinfo {volume} {635}},\ \bibinfo {pages} {584} (\bibinfo {year} {2024})}\BibitemShut {NoStop}%
\bibitem [{\citenamefont {Xia}\ \emph {et~al.}(2025)\citenamefont {Xia}, \citenamefont {Han}, \citenamefont {Watanabe}, \citenamefont {Taniguchi}, \citenamefont {Shan},\ and\ \citenamefont {Mak}}]{Xia2025Jan}%
  \BibitemOpen
  \bibfield  {author} {\bibinfo {author} {\bibfnamefont {Y.}~\bibnamefont {Xia}}, \bibinfo {author} {\bibfnamefont {Z.}~\bibnamefont {Han}}, \bibinfo {author} {\bibfnamefont {K.}~\bibnamefont {Watanabe}}, \bibinfo {author} {\bibfnamefont {T.}~\bibnamefont {Taniguchi}}, \bibinfo {author} {\bibfnamefont {J.}~\bibnamefont {Shan}},\ and\ \bibinfo {author} {\bibfnamefont {K.~F.}\ \bibnamefont {Mak}},\ }\bibfield  {title} {\bibinfo {title} {{Superconductivity in twisted bilayer WSe2}},\ }\href {https://doi.org/10.1038/s41586-024-08116-2} {\bibfield  {journal} {\bibinfo  {journal} {Nature}\ }\textbf {\bibinfo {volume} {637}},\ \bibinfo {pages} {833} (\bibinfo {year} {2025})}\BibitemShut {NoStop}%
\bibitem [{\citenamefont {Guo}\ \emph {et~al.}(2025)\citenamefont {Guo}, \citenamefont {Pack}, \citenamefont {Swann}, \citenamefont {Holtzman}, \citenamefont {Cothrine}, \citenamefont {Watanabe}, \citenamefont {Taniguchi}, \citenamefont {Mandrus}, \citenamefont {Barmak}, \citenamefont {Hone}, \citenamefont {Millis}, \citenamefont {Pasupathy},\ and\ \citenamefont {Dean}}]{Guo2025Jan}%
  \BibitemOpen
  \bibfield  {author} {\bibinfo {author} {\bibfnamefont {Y.}~\bibnamefont {Guo}}, \bibinfo {author} {\bibfnamefont {J.}~\bibnamefont {Pack}}, \bibinfo {author} {\bibfnamefont {J.}~\bibnamefont {Swann}}, \bibinfo {author} {\bibfnamefont {L.}~\bibnamefont {Holtzman}}, \bibinfo {author} {\bibfnamefont {M.}~\bibnamefont {Cothrine}}, \bibinfo {author} {\bibfnamefont {K.}~\bibnamefont {Watanabe}}, \bibinfo {author} {\bibfnamefont {T.}~\bibnamefont {Taniguchi}}, \bibinfo {author} {\bibfnamefont {D.~G.}\ \bibnamefont {Mandrus}}, \bibinfo {author} {\bibfnamefont {K.}~\bibnamefont {Barmak}}, \bibinfo {author} {\bibfnamefont {J.}~\bibnamefont {Hone}}, \bibinfo {author} {\bibfnamefont {A.~J.}\ \bibnamefont {Millis}}, \bibinfo {author} {\bibfnamefont {A.}~\bibnamefont {Pasupathy}},\ and\ \bibinfo {author} {\bibfnamefont {C.~R.}\ \bibnamefont {Dean}},\ }\bibfield  {title} {\bibinfo {title} {{Superconductivity in 5.0$\,^{\circ}$ twisted bilayer WSe$_2$}},\ }\href {https://doi.org/10.1038/s41586-024-08381-1} {\bibfield
  {journal} {\bibinfo  {journal} {Nature}\ }\textbf {\bibinfo {volume} {637}},\ \bibinfo {pages} {839} (\bibinfo {year} {2025})}\BibitemShut {NoStop}%
\bibitem [{\citenamefont {Geier}\ and\ \citenamefont {Nazaryan}(2025)}]{periodicwave_github}%
  \BibitemOpen
  \bibfield  {author} {\bibinfo {author} {\bibfnamefont {M.}~\bibnamefont {Geier}}\ and\ \bibinfo {author} {\bibfnamefont {K.}~\bibnamefont {Nazaryan}},\ }\href {http://github.com/mg607/periodicwave} {\bibinfo {title} {{PeriodicWave}}} (\bibinfo {year} {2025})\BibitemShut {NoStop}%
\bibitem [{\citenamefont {Cassella}\ \emph {et~al.}(2023{\natexlab{b}})\citenamefont {Cassella}, \citenamefont {Sutterud}, \citenamefont {Azadi}, \citenamefont {Drummond}, \citenamefont {Pfau}, \citenamefont {Spencer},\ and\ \citenamefont {Foulkes}}]{casella2023qpt}%
  \BibitemOpen
  \bibfield  {author} {\bibinfo {author} {\bibfnamefont {G.}~\bibnamefont {Cassella}}, \bibinfo {author} {\bibfnamefont {H.}~\bibnamefont {Sutterud}}, \bibinfo {author} {\bibfnamefont {S.}~\bibnamefont {Azadi}}, \bibinfo {author} {\bibfnamefont {N.~D.}\ \bibnamefont {Drummond}}, \bibinfo {author} {\bibfnamefont {D.}~\bibnamefont {Pfau}}, \bibinfo {author} {\bibfnamefont {J.~S.}\ \bibnamefont {Spencer}},\ and\ \bibinfo {author} {\bibfnamefont {W.~M.~C.}\ \bibnamefont {Foulkes}},\ }\bibfield  {title} {\bibinfo {title} {{Discovering Quantum Phase Transitions with Fermionic Neural Networks}},\ }\href {https://doi.org/10.1103/PhysRevLett.130.036401} {\bibfield  {journal} {\bibinfo  {journal} {Phys. Rev. Lett.}\ }\textbf {\bibinfo {volume} {130}},\ \bibinfo {pages} {036401} (\bibinfo {year} {2023}{\natexlab{b}})}\BibitemShut {NoStop}%
\end{thebibliography}%

\newpage
\appendix

\section{Neural architecture}


Here we describe the structure of the neural network from Fig. \ref{fig:scheme}(b) used to construct the generalized coordinates.

Starting with the state vector Eq. \eqref{eq:coordinates}, we transformed it into a real $d_{\rm feature}$-dimensional feature vector:
\bea
\bold{f}_i = \text{feature}(\bold{l}_i) , 
\eea
where the $\text{feature}$ function is problem specific, for example identity or a periodic function that enforces periodic boundary conditions \cite{pfau2020ab,von2022self,casella2023qpt,geier2025attention}. 

As the next step, the feature vector is embedded into internal representation as a real $d_L$-dimensional vector by a linear transformation:
\bea
\bold{h}^0_i = W^0 \bold{f}_i,
\eea
where $W^0 \in \mathbb{R}^{d_L} \times \mathbb{R}^{d_{\rm feature}}$. 
The same matrix $W^0$ is applied to the feature $\bold{f}_i$ from each electron.
All $\bold{h}^0_i$ 
This is followed by a sequence of self-attention and multi-layer perceptron layers (MLP), like in the Psiformer architecture 
\cite{pfau2020ab} based on original transformer 
\cite{vaswani2017attention}.

Self-attention is a way to include electron correlation. It is built out of “keys”, “queries”, and “values” computed for each electron stream:
\bea
\bold{k}^{lh}_i = W^{lh}_{\rm k} \bold{h}^l_i,~
\bold{q}^{lh}_i = W^{lh}_{\rm q} \bold{h}^l_i,~
\bold{v}^{lh}_i = W^{lh}_{\rm v} \bold{h}^l_i,
\eea
with $W^{lh}_{\rm k}, W^{lh}_{\rm q} \in \mathbb{R}^{d_L} \times \mathbb{R}^{d_{\rm Attn}}$ and $W^{lh}_{\rm v} \in \mathbb{R}^{d_L} \times \mathbb{R}^{d_{\rm AttnVals}}$. Here $l$ enumerates the layer and $h = 1, \dots, N_{\rm heads}$ enumerates the attention head (multiple attention calculation are performed in parallel).
The self-attention is then evaluated as 
\bea
{\rm \textsc{SelfAttn}}^{lh}_i = \frac{1}{\mathcal{N}} \sum_{j=1}^{N} \exp\left(\frac{\bold{q}_j^{lh}\cdot\bold{k}_i^{lh}}{\sqrt{d_{\rm Attn}}} \right)\bold{v}^{lh}_j,
\eea
where $\mathcal{N}$ is a normalization constant \cite{von2022self,geier2025attention}.
Now we update the embedded vectors:
\bea
\bold{f}^{l+1}_i = \bold{h}^{l}_i + W_o^l \text{concat}_h[{\rm \textsc{SelfAttn}}^{lh}_i],
\eea
with $W_o^l \in \mathbb{R}^{d_L} \times \mathbb{R}^{N_{\rm heads}d_{\rm AttnVals}}$.
After self-attention, the MLP is applied stream-wise:
\bea
\bold{h}^{l+1}_i = \bold{f}^{l+1}_i + \tanh(W^{l+1}\bold{f}^{l+1}+\bold{b}^{l+1}),
\eea
where $W^{l+1} \in \mathbb{R}^{d_L} \times \mathbb{R}^{d_L}$ and $\bold{b}^{l+1} \in \mathbb{R}^{d_L}$.
Next, we construct complex-valued generalized orbitals for each particle:
\bea\label{eq:gen-orbitals}
\phi^m_j(\bxi_i, \bxi_{/i}) = \bold{w}^m_{2j} \cdot \bold{h}^l_i + i~\bold{w}^m_{2j+1} \cdot \bold{h}^l_i,
\eea
where $\bold{w}^m_{2j}, \bold{w}^m_{2j+1}$ are (real) learnable projection vectors. Finally, we take a sum determinant of the generalized coordinates, Eq. \eqref{eq:generalized_determinant_wf}, to obtain the wavefunction that obeys the Pauli principle. Note, that the orbitals do not need to be orthogonal.

\section{Simulation hyperparameters} \label{app:hyperparameters}
We report the simulations’ hyperparameters (Sec.~\ref{sec:results}) in Table~\ref{tab:hyperparams}. All three experiments used a $\eta_{0}\!\left(1+\frac{t}{t_{0}}\right)^{-1}$ scheduler.
\begin{table}[t]
\centering
\caption{The hyperparameters used in our numerical calculations. }
\label{tab:hyperparams}
\begin{tabular}{llll}
\hline
\textbf{Parameter} & \textbf{Sec.} \ref{sec:spin-spiral} & \textbf{Sec.} \ref{sec:soc} & \textbf{Sec.} \ref{sec:afm} \\
\hline
\multicolumn{4}{l}{\textbf{Architecture}} \\
Network layers & 4 & 4 & 2 \\
Attention heads per layer & 4 & 4 & 4 \\
Attention dimension & 16 & 16 & 32\\
Perceptron dimension & 64 & 64 & 128 \\
\# perceptrons per layer & 1 & 1 & 2\\
Determinants & 4 & 4 & 4 \\
\hline
\multicolumn{4}{l}{\textbf{Training}} \\
Training iterations & $2\times10^{4}$ & $2.5\times10^{3}$ & $1.5\times10^{5}$ \\
Initial learning rate $\eta_{0}$ & $0.01$ & $0.02$ & 0.01 \\
Learning rate delay $t_{0}$ & $10^{5}$ & $10^{5}$ & $2 \times 10^{5}$ \\
Local energy clipping $\rho$ & $5.0$ & $5.0$ & $5.0$\\
\hline
\multicolumn{4}{l}{\textbf{MCMC}} \\
Batch size & 1024 & 2048 & 1024 \\
\hline
\multicolumn{4}{l}{\textbf{KFAC}} \\
Norm constraint & $10^{-3}$ & $10^{-3}$ & $10^{-3}$ \\
Damping & $10^{-3}$ & $10^{-3}$ & $10^{-4}$ \\
\hline
\end{tabular}
\end{table}

\end{document}